\documentclass[preprints,article,accept,oneauthor,pdftex]{Definitions/mdpi} 

\usetikzlibrary{decorations.pathreplacing,decorations.markings}
\usetikzlibrary{decorations.pathmorphing}
\usetikzlibrary{shapes.geometric,positioning}

\tikzset{
operator/.style = {shape = circle, thick, draw, fill, minimum size = 1.5ex,inner sep=0,outer sep=0},
altoperator/.style = {shape = rectangle, thick, draw, fill, minimum size = 1.5ex,inner sep=0,outer sep=0},
fermion/.style = {thick, postaction={decorate}, decoration={markings,mark=at position 0.6 with {\arrow{latex}}}},
boson/.style = {very thick, dashed},
gaugeboson/.style = {very thick, decorate,decoration={snake}}
}

\firstpage{1} 
\pubvolume{1}
\issuenum{1}
\articlenumber{0}
\pubyear{2021}
\copyrightyear{2020}
\datereceived{} 
\dateaccepted{} 
\datepublished{} 
\hreflink{https://doi.org/} 

\Title{Beta Decay in Medium-Mass Nuclei with the In-Medium Similarity Renormalization Group}

\TitleCitation{Title}

\Author{S Ragnar Stroberg $^{1,\dagger,\ddagger}$\orcidA{}}

\AuthorNames{S Ragnar Stroberg}

\AuthorCitation{Stroberg, S. R.}

\address{%
$^{1}$ \quad Physics Division, Argonne National Laboratory, Lemont, IL, 60439, USA; sstroberg@anl.gov}

\corres{Correspondence: sstroberg@anl.gov}

\abstract{We review the status of ab initio calculations of allowed beta decays (both Fermi and Gamow-Teller), within the framework of the valence-space in-medium similarity renormalization group approach.}


\begin{document}
\section{Introduction}
Beta decays in atomic nuclei have long been a source of fundamental discoveries in physics~\cite{Fermi1934,Lee1956,Wu1957}, and precise measurements of beta decays continue to be a promising path to search for physics beyond the Standard Model (BSM)~\cite{Avignone2008,Holstein2014,Hardy2015,Hayen2018}.
A major challenge in the search for a signal of new physics is understanding the Standard Model ``background'', especially the effects of low-energy quantum chromodynamics which manifest as nuclear structure.
The situation is aggravated by the fact that nuclei which are preferred experimentally are often difficult to treat theoretically in a framework that allows quantified uncertainties.

Nevertheless, progress has been made over the past few decades so that the internucleon interaction can be systematically constructed within an effective field theory framework~\cite{Epelbaum2009,Machleidt2011,Hammer2019}.
Simultaneously, advances in many-body theory and computational resources have enabled ab initio treatment of the medium-mass nuclei which are often relevant for BSM searches~\cite{Carlson2015,Hagen2014,Binder2014,Morris2018,Hergert2020,Soma2020a,Lee2020}.
Of course more work remains to be done, both on the effective field theory side~\cite{VanKolck2020,Phillips2021,Cirigliano2021}, and on understanding how approximation schemes in ab initio calculations impact the observables in question.

In this paper, I will focus on one particular many-body method---the valence-space in-medium similarity renormalization group (VS-IMSRG)---and consider two topics in allowed beta decay, the quenching in Gamow-Teller decays and correction factors for superallowed $0^+\to 0^+$ Fermi decays.

\section{IMSRG formalism}
There are several review articles detailing both the free-space SRG~\cite{Bogner2010,Furnstahl2013} and the in-medium SRG~\cite{Hergert2016,Hergert2017,Hergert2017a,Stroberg2019,Hergert2020}, and so here I will review only what is needed for our present purposes.

\subsection{Similarity renormalization group}
The basic idea of the SRG is to perform a unitary transformation $U$ on the Hamiltonian $H$ (and all other operators) in such a way that the resulting nuclear wave function is simpler.
This is achieved by performing a sequence of infinitessimal unitary transformations, labeled by a flow parameter $s$, so that
\begin{equation}\label{eq:Hs}
    H(s) = U(s) H(0) U^{\dagger}(s)
\end{equation}
with $U(0)=1$.
The way in which $U$ changes with $s$ is determined by the generator $\eta$, which we are free to specify,
\begin{equation}\label{eq:dUds}
    \frac{dU(s)}{ds} = \eta(s) U(s)
\end{equation}
as long as $\eta$ is anti-hermitian $\eta^{\dagger}=-\eta$.
Combining \eqref{eq:Hs} and \eqref{eq:dUds} we obtain a flow equation for the Hamiltonian in terms of a commutator with the generator
\begin{equation}\label{eq:dHds}
    \frac{dH}{ds} = [\eta(s),H(s)].
\end{equation}
The flow equation for any other operator $\mathcal{O}$ is obtained by replacing $H\to \mathcal{O}$ in \eqref{eq:dHds}~\cite{Anderson2010,Schuster2014,Parzuchowski2017,Tropiano2020}.

It remains to specify $\eta(s)$.
In the free-space SRG, we choose
\begin{equation}
    \eta^{SRG}(s) = [T,V(s)]
\end{equation}
where $T$ is the kinetic energy and $V(s)$ is the potential so $H(s)=T+V(s)$.
This generator drives $V(s)$ towards a band-diagonal form in momentum space, with a width $\lambda_{SRG}\equiv s^{-1/4}$.

When the SRG flow equation \eqref{eq:dHds} is formulated in Fock-space (i.e. in terms of creation and annihilation operators), many-body forces are inevitably induced, and these must be truncated in order to make the calculation tractable.
For this reason, the ``free-space'' SRG evolution is typically performed out to $\lambda_{SRG}\gtrsim 2{\rm fm}^{-1}$.

\subsection{In-medium SRG}
The truncation of many-body forces is rendered less severe if all operators are normal-ordered with respect to a reference state $|\Phi\rangle$, which should be a reasonable first approximation of the exact wave function $|\Psi\rangle$.
This approach is called the in-medium SRG (IMSRG).
If we choose the generator to suppress the parts of $H$ which lead to excitations out of the reference $|\Phi\rangle$, then for $s\to \infty$ $|\Phi\rangle$ becomes an eigenstate of $H(s)$ with an eigenvalue corresponding to the energy of the exact wave function $|\Psi\rangle$, up to approximation errors in solving \eqref{eq:dHds}.
In all calculations presented here, I neglect three-body operators after the initial normal-ordering step, resulting in the IMSRG(2) approximation.

One possible choice for the generator which achieves the desired suppression was proposed by White~\cite{White2002}
\begin{equation}\label{eq:WhiteGen}
    \eta^{\rm Wh} \equiv \frac{H^{\rm od}}{\Delta}
\end{equation}
where the ``off-diagonal'' part of the Hamiltonian, denoted $H^{\rm od}$, is any part of $H$ which connects $|\Phi\rangle$ to a different state.
The energy denominator $\Delta$ is typically defined with Epstein-Nesbet or M{\o}ller-Plesset partitioning.
In this work, I use Epstein-Nesbet denominators, and a modification of \eqref{eq:WhiteGen}---also suggested by White---called the arctangent generator (see Ref.~\cite{Stroberg2019} for more details).

As a further generalization, we may define a valence space (e.g. the sd-shell above an $^{16}$O core), and define $H^{\rm od}$ such that any part of $H$ which connects a valence configuration to a non-valence configuration is suppressed\footnote{Specifically, we partition all single-particle states into core, valence, and excluded orbits.  A ``valence configuration'' is one with all core orbits occupied and all excluded orbits unoccupied.}.
The Hamiltonian is then driven to a block-diagonal form and we may diagonalize in the (typically much smaller) sub-space of valence configurations.
Such a diagonalization directly corresponds to a standard large-scale shell model calculation with an effective interaction defined by $H(\infty)$.
This approach is referred to as the valence-space IMSRG (VS-IMSRG), and it is used for all calculations presented.

Generally, the states we wish to target in a valence space approach are not well-described by a single closed-shell configuration, and the choice of $|\Phi\rangle$ becomes less clear.
In this work I use the ensemble normal-ordering (ENO) approach~\cite{Stroberg2017}, which amounts to taking fractional occupation numbers such that the reference has spherical symmetry and the correct number of particles on average.
In addition, I use the Magnus formulation of the IMSRG~\cite{Morris2015}, in which we write
\begin{equation}\label{eq:MagnusU}
 U(s) \equiv e^{\Omega(s)}
\end{equation}
where $\Omega=-\Omega^\dagger$ is the Magnus operator.
Equations \eqref{eq:MagnusU} and \eqref{eq:dUds} may be combined to obtain a flow equation for $\Omega(s)$ in terms of $\eta(s)$, and operators (including the Hamiltonian), are transformed as
\begin{equation}\label{eq:MagnusBCH}
\begin{aligned}
    \mathcal{O}(s) &= e^{\Omega(s)}\mathcal{O}(0)e^{-\Omega(s)} \\
    &=\mathcal{O}(0) + [\Omega(s),\mathcal{O}(0)] + \tfrac{1}{2!}\bigl[\Omega(s),[\Omega(s),\mathcal{O}(0)]\bigr]+\ldots
    \end{aligned}
\end{equation}
In \eqref{eq:MagnusBCH}, each commutator is truncated at the normal-ordered two-body level, and the series is computed iteratively until the size of a term falls below a numerical threshold.

\subsection{Aspects relevant to beta decay}
Two additional details of the calculation are relevant for $\beta$ decays.
The first is the choice of single-particle basis in which we express the operators at $s=0$.
In this work I use a Hartree-Fock basis with Coulomb and isospin-breaking strong forces included, so that for a given set of single-particle quantum numbers $\{n,\ell,j\}$, the proton and neutron radial wave functions are not identical.
The second is the choice of reference.
Because we wish to compute the initial state, with $N$ neutrons and $Z$ protons, consistently with the final state with $N\pm 1$ neutrons and $Z\mp 1$ protons, there is some ambiguity about which reference $|\Phi\rangle$ should be used.
If we retain all induced many-body terms during the SRG evolution, the choice of reference is irrelevant.
However, the accuracy of the IMSRG(2) approximation depends on the choice of reference.
The two natural choices for $\beta$ decay are to use the $N,Z$ of the initial state or the final state.
I will discuss this in more detail in Section~\ref{sec:Fermi}.

\section{Gamow-Teller decays\label{sec:GT}}
In a Gamow-Teller decay, the leptons carry one unit of angular momentum and leave the parity of the nucleus unchanged.
The relevant nuclear transition operator is obtained from the space-like part of the hadronic axial-vector current.
The leading term in the non-relativistic reduction is $g_A \sigma\tau$, where $g_A\approx 1.27$ is the axial coupling constant, and $\sigma$ and $\tau$ are the spin and isospin Pauli matrices.

Historically, when the leading operator was combined with shell model wave functions, a systematic ``quenching'' of the decay strength was observed, i.e. experimental matrix elements were smaller than the predicted ones, with a similar effect in isovector $M1$ observables~\cite{Wilkinson1973a,Wilkinson1973,Brown1978,Brown1985a,Chou1993,Martinez-Pinedo1996}.
It was quickly surmised that the source of the discrepancy should be the some combination of inadequate wave functions (missing correlations) and an inadequate transition operator (missing currents), that neither of these obviously dominated, and that the two effects were not independent~\cite{Wilkinson1973,Rho1974,Towner1983}.
It was also suggested that pions ought to have something to do with the renormalization of the axial current in the nuclear medium~\cite{Ericson1973}.

These physics arguments survive in the modern EFT point of view, which organizes the nuclear interaction and coupling to external fields in powers of a ratio of scales.
The distinction between short and long-distance physics is made by a cutoff, and the arbitrariness of the cutoff is reflected in the requirement that observables be independent of its value.
The relationship between pions and the axial current arises as a consequence of broken chiral symmetry~\cite{Scherer2012}.
Importantly, chiral EFT enables a systematic and consistent construction of three-nucleon forces and two-body currents~\cite{Park2003,Gardestig2006,Gazit2009,Krebs2017}.

\begin{figure}
    \centering
    \begin{tikzpicture}
\draw [fermion] (-0.5,-1) -- (-0.5,0);
\draw [fermion] (-0.5,0) -- (-0.5,1);
\draw [gaugeboson] (-1.5,-0.0)node{\Large$\times$} -- (-0.5,-0.0);
\node (abc) at (-0.8,-1.7) {(a)};
\end{tikzpicture}
    \hspace{2em}
    \begin{tikzpicture}
\draw [fermion] (-0.5,-1) -- (-0,0);
\draw [fermion] (0,0) -- (-0.5,1);
\draw [fermion] (0.5,-1) -- (0.0,0);
\draw [fermion] (0.0,0) -- (0.5,1);
\draw [gaugeboson] (-1.0,-0.0)node{\Large$\times$} -- (0.0,-0.0);
\fill[black] (-0.0,0) circle (0.5ex);
\node (abc) at (-0.5,-1.7) {(b)};
\end{tikzpicture}
    \hspace{2em}
    \begin{tikzpicture}
\draw [fermion] (-0.5,-1) -- (-0.5,0);
\draw [fermion] (-0.5,0) -- (-0.5,1);
\draw [fermion] (0.5,-1) -- (0.5,0);
\draw [fermion] (0.5,0) -- (0.5,1);
\draw [gaugeboson] (-1.4,-0.0)node{\Large$\times$} -- (-0.5,-0.0);
\draw [boson] (-0.5,0) --node[above]{$\pi$} (0.5,0);
\fill[black] (-0.5,0) circle (0.5ex);
\node (abc) at (-0.5,-1.7) {(c)};
\end{tikzpicture}
    \caption{Diagrams for (a) leading-order Gamow-Teller decay $\sigma\tau$, (b) short-range two-body current, and (3) long-range two-body current.}
    \label{fig:axialDiagrams}
\end{figure}
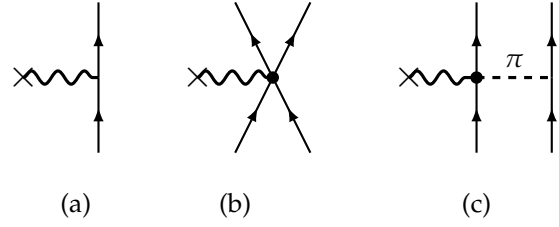

The result in the limit that the momentum carried by the leptons vanishes, up to order $Q^0$ (leading order is $Q^{-3}$) is~\cite{Park1997,Park2003,Menendez2011}
\begin{equation}
    \vec{J} = \vec{J}_{1b} + \vec{J}_{\rm 2b; cont} + \vec{J}_{{\rm 2b;}1\pi}
\end{equation}
where
\begin{equation}\label{eq:J1b}
    \vec{J}_{1b}^{\pm} = g_A \vec{\sigma}\tau^{\pm}
\end{equation}
\begin{equation}\label{eq:J2bcont}
    \vec{J}_{\rm 2b; cont}^{\pm} = \frac{1}{2} \frac{c_D}{\Lambda_{\chi}f_{\pi}^2} (\vec{\sigma}_1 \tau_1^{\pm} +\vec{\sigma}_2 \tau_2^{\pm})   
\end{equation}
\begin{equation}\label{eq:J2bpi}
    \vec{J}_{{\rm 2b;} 1\pi} = 
    -\frac{g_A}{f_\pi^2 }
 \frac{\vec{\sigma}_2\cdot \vec{q}_2}{k_2^2 + M_\pi^2}
 \left[\frac{i\vec{p_1}}{2m}\tau_\times^{\pm}
  + 2c_3 \tau_2^{\pm} \vec{k}_2
    + (c_4+\frac{1}{4m}) \tau_\times^{\pm} (\vec{\sigma}_1\times\vec{k}_2)
    \right]
    +(1\leftrightarrow 2)
\end{equation}
where $\vec{p_i}$,$\vec{p'}_i$ are the incoming and outgoing momenta of the $i$th nucleon, $\vec{k}_i=\vec{p'}_i-\vec{p}_i$, $\tau_\times = \tau_1\times \tau_2$, and $f_\pi$ is the pion decay constant.
The low-energy constants $c_3$, $c_4$ and $c_D$ also enter into the NN and 3N forces, and so are not additional free parameters.
Equations \eqref{eq:J1b}, \eqref{eq:J2bcont}, and \eqref{eq:J2bpi} correspond to diagrams (a), (b), and (c) in Figure~\ref{fig:axialDiagrams}, respectively.
Note that there are also corrections to the one-body operator of order $p^2_i/m_N^2$.
Depending on how the nucleon mass is counted, these corrections will enter at different orders.
In the counting of e.g. Park et al~\cite{Park2003}, these corrections are also $Q^0$, while in the counting used by other authors~\cite{Menendez2011,Gysbers2019,Krebs2017}, including the calculations in this paper, these corrections are $Q^1$.

In Refs.~\cite{Park1997} and ~\cite{Menendez2011}, these currents were normal ordered with respect to uniform nuclear matter to obtain an in-medium quenching factor for the one-body operator.
In Ref.~\cite{Ekstrom2014}, the full two-body current was constructed, consistently\footnote{In \cite{Ekstrom2014}, the relationship between the two-body currents and three-body force contained an erroneous factor of $-1/4$~\cite{Gazit2019}.} with the NN+3N force, and Gamow-Teller decays of $^{14}$C, $^{22}$O and $^{24}$O were computed using the coupled cluster method~\footnote{I also note that while the decay of $^{14}$C is interesting due to the anomolously long half-life~\cite{Holt2008,Maris2011}, the small matrix element makes it difficult to draw conclusions regarding systematic quenching effects.}.
In all three of these cases, a quenching of about the right size was obtained.
In Ref~\cite{Pastore2018}, axial currents up to N$^4$LO were used in quantum Monte Carlo calculations of $A$=6-10 nuclei, where it was found that correlations beyond the shell model accounted for most of the quenching, with subleading currents playing a minor role.
In Ref.~\cite{Gysbers2019}, the full two-body current up to N$^3$LO was constructed consistently with the NN+3N force, consistently SRG evolved, and evaluated in a range of nuclei in the $p$, $sd$, and $pf$ shells, as well as $^{100}$Sn, using no-core shell model, coupled cluster, or VS-IMSRG to solve the many-body problem.
Here, I will provide some additional calculations not presented in \cite{Gysbers2019}, and some further discussion.

The experimental Gamow-Teller matrix elements are obtained from the $ft$ values by
\begin{equation}\label{eq:ft}
    ft = \frac{K}{\left[\tfrac{f_V}{f_A}B(F) + B(GT)\right]G_V^2}
\end{equation}
with $K\equiv (2\pi^3 \hbar^7\ln 2)/ (m_e^5c^4)$, and $K/G_V^2\approx 6140$s.
The Gamow-Teller matrix element is defined as
\begin{equation}\label{eq:MGT}
    M(GT) \equiv \left[ (2J_i+1)B(GT)\right]^{1/2}.
\end{equation}
Note different definitions have been used in the literature, e.g. one may divide the right hand side by $g_A$, as was done in~\cite{Gysbers2019}.

I consider Gamow-Teller transitions in nuclei in the $p$, $sd$ and $pf$ shells, with experimental data taken from refs.~\cite{Chou1993,Brown1985a,Martinez-Pinedo1996}.
I have selected transitions with large transition matrix elements, with the goal of reducing sensitivity to fine-tuned cancellations.
I also consider the decay of $^{100}$Sn, which was treated with equations-of-motion coupled cluster in~\cite{Gysbers2019}, and for which the experimental picture is still somewhat conflicted~\cite{Faestermann2002,Batist2010,Hinke2012,Lubos2019}.
I adopt the average value presented in~\cite{Lubos2019}.
In the VS-IMSRG calculation of $^{100}$Sn, I use valence space consisting of the $0f_5,1p_3,1p_3,0g_9$ orbits for protons and $0f_7,1d_5,1d_3,2s_1,0h_{11}$ for neutrons.

For the theoretical calculations, I use the NN+3N(lnl) interaction developed by Navr\'atil~\cite{Soma2020}.
The interaction and current are consistently SRG-evolved to a scale $\lambda_{SRG}=2.0~{\rm fm}^{-1}$ and evaluated in an oscillator space defined by $2n+\ell\leq e_{\rm max}=12$ and $\hbar\omega=16$.
The 3N matrix elements are further truncated with $e_1+e_2+e_3\leq E_{\rm 3max}=14$.
All operators are transformed to the Hartree-Fock basis, and then the residual 3N operators are truncated (the NO2B approximation).
Next, a VS-IMSRG calculation is performed using the code \texttt{imsrg++}\cite{imsrgcode}, yielding and effective valence space interaction and operator.
The valence space diagonalization is carried out either using \texttt{NuShellX@MSU}~\cite{Brown2014} with operators evaluating using the code \texttt{nutbar}\cite{nutbarcode}, or with \texttt{KSHELL}~\cite{Shimizu2013,Shimizu2019}.
The results are listed in Table~\ref{tab:GTresults} and plotted in Fig.~\ref{fig:GT2panel}.

Table~\ref{tab:GTresults} in the appendix contains the numerical results.
The column labeled $M(GT)_{\rm exp}$ lists the experimental Gamow-Teller matrix elements defined by~\eqref{eq:MGT} (experimental uncertainties are not listed).
The column labeled $\sigma\tau_{\rm bare}$ is the obtained by evaluating the operator $\sigma\tau$ (assuming identical radial wave functions for protons and neutrons) between valence space wave functions obtained using the VS-IMSRG evolved interaction.
The column labeled $\sigma\tau_{\rm IMSRG}$ is obtained by consistently SRG and VS-IMSRG evolving the $\sigma\tau$ operator (including the radial mismatch due to the Hartree-Fock basis).
Finally, $M(GT)_{\rm th}$ also includes the two-body currents, consistently SRG and VS-IMSRG evolved.
In a few cases, the listed strength is summed over multiple final states with the same spin and parity.

In Fig.~\ref{fig:GT2panel}, panel (a) shows a scatter plot of $M(GT)_{\rm exp}$ vs $M(GT)_{\rm bare}$, while panel (b) shows $MGT)_{\rm th}$ vs $M(GT)_{\rm exp}$.
The solid line shows $y=x$ corresponding to perfect agreement between theory and experiment.
The dashed line shows a best-fit slope, which is indicated as a quenching factor at the top of the figure.
For this quenching factor, I only include $sd$ and $pf$ shell nuclei because the $p$ shell nuclei have a large scatter due to nuclear structure details.
The quantity in parenthesis indicates the standard deviation about the best-fit line.
If I include $p$ shell nuclei in the fit, the full theory quenching factor changes to $q=0.99$, but the standard deviation increases to 0.21.

\begin{figure}[ht]
    \centering
    \includegraphics[width=\columnwidth]{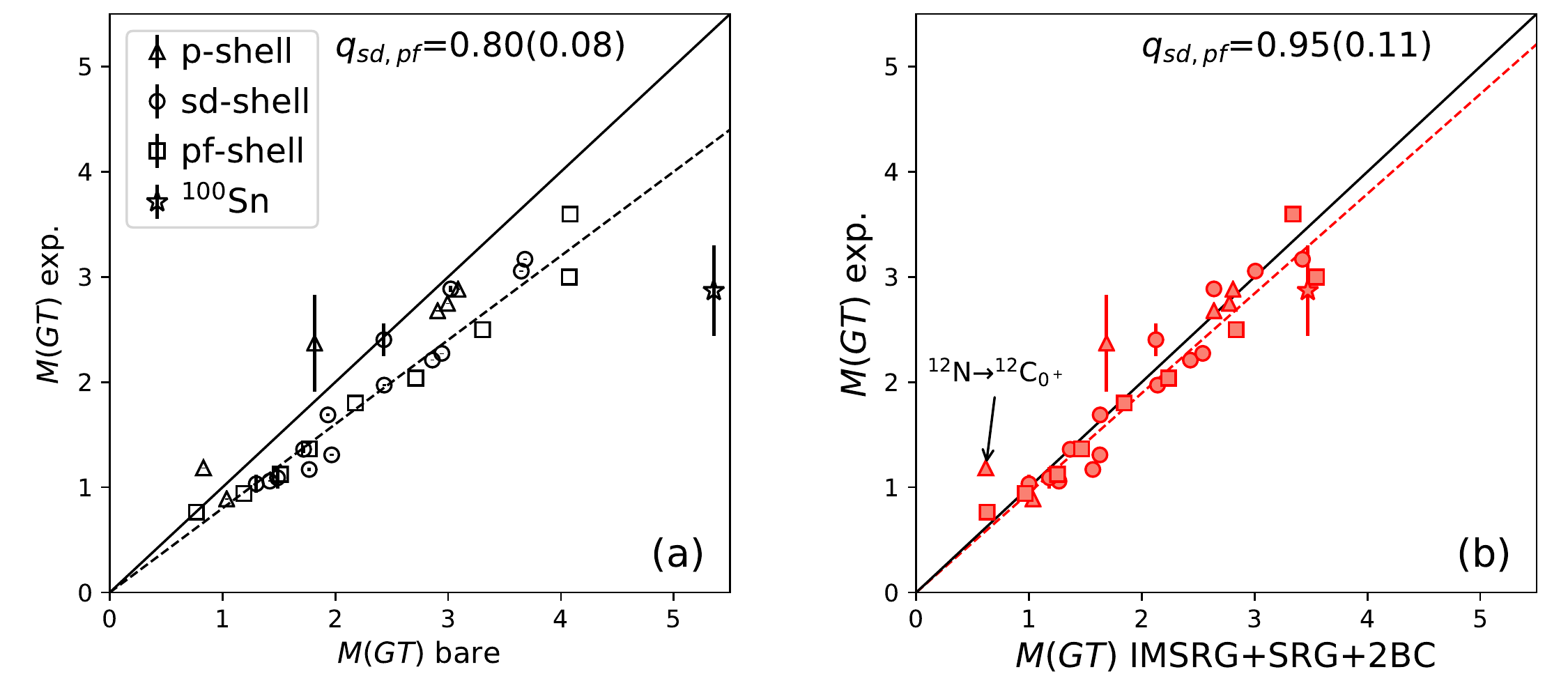}
    \caption{Experinmental $M(GT)$ vs $M(GT$ obtained with (a) the bare $g_A \sigma\tau$ operator, (b) the SRG- and IMSRG-transformed transition operator including two-body currents. In both panels, the solid line shows $y=x$ corresponding to perfect agreement, while the dashed line indicates the best-fit slope.}
    \label{fig:GT2panel}
\end{figure}

It is evident from Table~\ref{tab:GTresults} that both the correlations included in $\sigma\tau_{\rm IMSRG}$ and the two-body currents lead to a reduction of the Gamow-Teller matrix element.
As discussed in Ref.~\cite{Gysbers2019}, the detailed breakdown of the quenching into correlations and currents is scheme- and scale-dependent; some of the effects attributed to correlations when using a hard interaction get shuffled into currents, when using a soft interaction.

It is also evident from Figure~\ref{fig:GT2panel} that the systematic quenching effect, observed when using the bare $\sigma\tau$ operator, essentially vanishes when using the consistently evolved operator including two-body currents.

In the right panel of Figure~\ref{fig:GT2panel} I highlight the transition $^{12}$N$\to^{12}$C$_{0^+}$ as an illustration of the cancellation effects in the $p$ shell which wash out the quenching signal.
When evaluating the bare $\sigma\tau$ operator\footnote{The matrix element of the bare operator is not an observable, so strictly there's no reason different Hamiltonians should agree on it.
On the other hand one might expect some degree of universality within a low-resolution picture like the shell model~\cite{Bogner2003,Bogner2010}.} (including $g_A$), there are four terms that contribute, corresponding to proton-to-neutron transitions
$p_3 \to p_3$, $p_3 \to p_1$, $p_1 \to p_3$, and $p_1 \to p_1$.
The contributions are +0.366, -0.955, +1.592, and -0.173, respectively, totalling to 0.830.
Evidently there is significant cancellation so that a relatively small change of the individual terms can lead to a relatively large change on the final matrix element.
For example, if I use valence-space wave functions obtained with the phenomenological Cohen-Kurath interaction~\cite{Cohen1965}, the bare operator yields a matrix element of 1.219.
This difference, due to configuration mixing within the valence space, is larger than the systematic quenching effect of interest.

\begin{figure}[ht]
    \centering
    \includegraphics[width=\columnwidth]{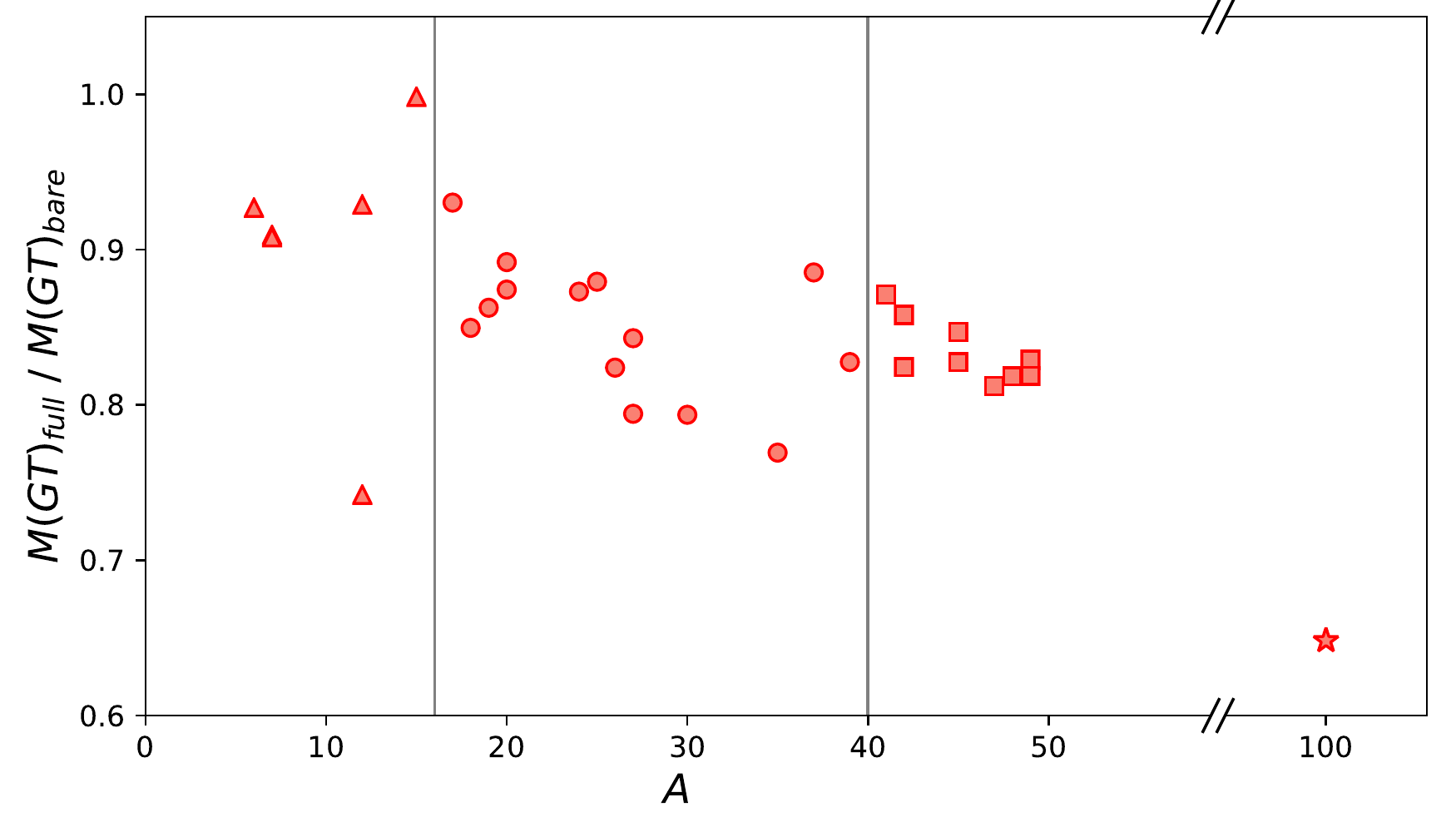}
    \caption{Theory-to-theory ``quenching'' factors as a function of mass number$A$.}
    \label{fig:GTq_vs_A}
\end{figure}

We may view this ``noise'' in the quenching from another perspective.
Fig.~\ref{fig:GTq_vs_A} shows the theory-to-theory quenching factor $M(GT)_{th} / M(GT)_{\rm bare}$ as a function of mass number.
This indicates the quenching factor needed if we wanted to approximately account for the correlations and currents contained in $M(GT)_{\rm th}$.
The main point here is to emphasize that the quenching is not a smooth function of $A$, but in fact has considerable state dependence.

Moving forward, including two-body currents consistently with the NN+3N interaction helps to remove the ambiguity of emprical quenching factors, especially in the context of double beta decay~\cite{Engel2017}.
Of course, sensitivity to details of the configuration mixing must be checked.

\section{Superallowed $0^+\to 0^+$ Fermi decays\label{sec:Fermi}}

For transitions between $J =0$ states, $B(GT)=0$ by conservation of angular momentum.
Furthermore, in the limit in which isospin is a perfect symmetry, a ``superallowed'' transition between $T=1$ isobaric analogue states yields $B(F)=2$, and so \eqref{eq:ft} reduces to\footnote{I am also neglecting here radiative corrections, which have a non-negligible impact.}
\begin{equation}\label{eq:ftIsospin}
    ft = \frac{K}{2G_V^2}
    \hspace{2em}
    {\rm(isospin~limit)}
\end{equation}
(where here $f=f_V)$.
This implies all superallowed $0^+\to 0^+$ should have the same $ft$ value, and that from this one may measure the coupling constant for semileptonic decay $G_V$, which is in turn related to the constant $G_F$ obtained from muon decay by $G_V=V_{ud}G_F$, where $V_{ud}$ is the up-down element of the Cabibbo-Kobayashi-Maskawa (CKM) quark mixing matrix.
Consequently, precise $ft$ measurements of superallowed $0^+\to 0^+$ decays provide a sensitive test of the Standard Model:
non-universality of superallowed $ft$ values, or non-unitarity of the CKM matrix would be signs of new physics.

Of course, isospin is not a perfect symmetry of the Standard Model.
It is broken by the quark electric charges, and the up-down mass difference.
This is manifested at the nuclear level as the Coulomb force between protons and isospin-violating strong interactions.
The Standard Model corrections to  \eqref{eq:ftIsospin} have been parameterized by Towner and Hardy~\cite{Hardy2015} as
\begin{equation}\label{eq:FtTH}
    \mathcal{F}t \equiv
    ft(1+\delta_R')(1+\delta_{\rm NS}-\delta_C) = \frac{K}{2G_V^2\Delta_R^V}.
\end{equation}
In \eqref{eq:FtTH} $\Delta_R^V$ is a process-independent radiative correction~\cite{Seng2019}, $\delta_R'$ is a radiative correction only depending on the electron energy and the charge of the daughter nucleus, and $\delta_{\rm NS}$ is a radiative correction depending on the detailed nuclear structure.
The isospin-symmetry-breaking correction $\delta_C$ accounts for the fact that the final state is not exactly an isospin rotation of the initial state.

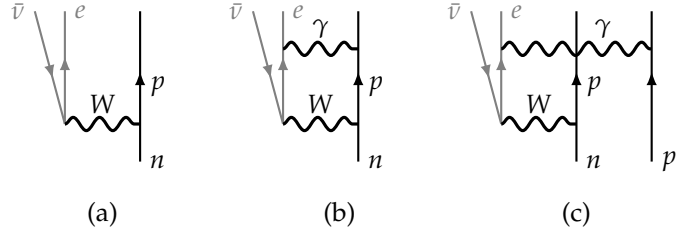
\begin{figure}
    \centering
    \begin{tikzpicture}
\draw [fermion,gray] (-1.5,-0.5) -- (-1.5,1)node[right]{$e$};
\draw [fermion,gray] (-1.9,1)node[left]{$\bar{\nu}$} -- (-1.5,-0.5);
\draw [fermion] (-0.5,-1)node[right]{$n$} --node[right]{$p$} (-0.5,1);
\draw [gaugeboson] (-1.5,-0.5) --node[above]{$W$} (-0.5,-0.5);
\node (abc) at (-1,-1.7) {(a)};
\end{tikzpicture}
    \hspace{1em}
    \begin{tikzpicture}
\draw [fermion,gray] (-1.5,-0.5) -- (-1.5,1)node[right]{$e$};
\draw [fermion,gray] (-1.9,1)node[left]{$\bar{\nu}$} -- (-1.5,-0.5);
\draw [fermion] (-0.5,-1)node[right]{$n$} --node[right]{$p$} (-0.5,1);
\draw [gaugeboson] (-1.5,0.5) --node [above]{$\gamma$} (-0.5,0.5);
\draw [gaugeboson] (-1.5,-0.5) --node[above]{$W$} (-0.5,-0.5);
\node (abc) at (-0.75,-1.7) {(b)};
\end{tikzpicture}
    \hspace{1em}
    \begin{tikzpicture}
\draw [fermion,gray] (-1.5,-0.5) -- (-1.5,1)node[right]{$e$};
\draw [fermion,gray] (-1.9,1)node[left]{$\bar{\nu}$} -- (-1.5,-0.5);
\draw [fermion] (-0.5,-1)node[right]{$n$} --node[right]{$p$} (-0.5,1);
\draw [fermion] (0.5,-1)node[right]{$p$} -- (0.5,1);
\draw [gaugeboson] (-1.5,0.5) --node [above]{\hspace{2em}$\gamma$} (0.5,0.5);
\draw [gaugeboson] (-1.5,-0.5) --node[above]{$W$} (-0.5,-0.5);
\node (abc) at (-0.5,-1.7) {(c)};
\end{tikzpicture}
    \caption{Schematic high-resolution diagrams corresponding to (a) the leading operator $\tau$, (b) the one-body correction leading to $\Delta_R^V$ and $\delta_R'$, and (c) the two-body correction leading to $\delta_{NS}$.}
    \label{fig:Superalloweddiagrams}
\end{figure}

Consequently, only $\delta_{\rm NS}$ and $\delta_{C}$ are the purview of nuclear structure theory.
To draw an analogy with the situation for Gamow-Teller decays, $\delta_{C}$ corresponds to including the effects of correlations for the leading operator $\tau$, while the radiative corrections correspond to sub-leading corrections to the operator, with $\delta_{\rm NS}$ corresponding to two-body currents.
The difference here is that the corrections are sub-leading in the fine structure constant $\alpha\approx 1/137$ (or $Z\alpha$), as opposed to the chiral EFT expansion parameter $Q\sim 1/4$.
The various corrections are illustrated in Fig.~\ref{fig:Superalloweddiagrams}.

In this paper we focus on the $\delta_C$ correction, for no better reason than the operator is the simplest to implement.
Towner and Hardy decompose $\delta_C$ into a correction due to isospin-breaking configuration mixing effects, and a correction due the mismatch in single-particle wave functions between protons and neutrons.
As we will be treating both within a consistent calculation, such  a decomposition is not necessary (and ambiguous) and we will simply use
\begin{equation}
    \delta_C = 1 - |M_F|^2/2
\end{equation}
where $M_F=\langle \Psi_f \| \tau \| \Psi_i\rangle$ is the result of the ab initio calculation.

Nevertheless, it is useful to keep the two mechanisms (configuration mixing and wave function mismatch) in mind when considering the impact of various approximations.
The wave function mismatch effect is taken into account primarily by the fact that we use a Hartree-Fock single-particle basis, with Coulomb and nuclear ISB effects included in the potential (see also~\cite{Miller2009}).

\begin{figure}
    \centering
    \includegraphics[width=\columnwidth]{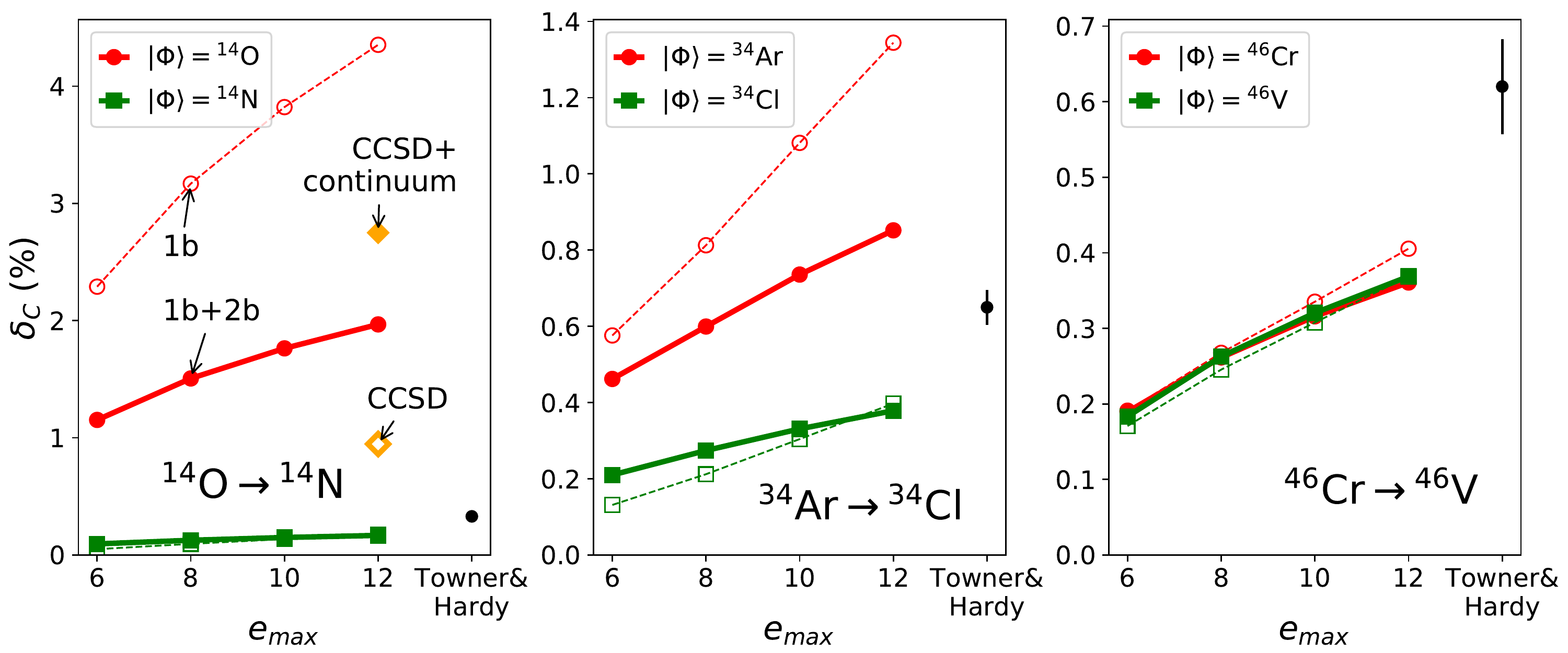}
    \caption{Isospin-breaking correction $\delta_C$ for three transitions, computed with the VS-IMSRG as a function of the $e_{\rm max}$ truncation. Dashed lines show the result using only the one-body part of the evolved operator while solid lines also include induced two-body terms. The black circles indicate the preferred values reported by Towner and Hardy~\cite{Hardy2015}. For the $^{14}$O decay, we also include coupled cluster points~\cite{HagenPC}.}
    \label{fig:deltaC3panel}
\end{figure}

To get an idea of what the VS-IMSRG framework produces for the isospin-breaking correction $\delta_C$, I consider three transitions spanning the $p$, $sd$ and $fp$ shells: $^{14}$O$\to ^{14}$N, $^{34}$Ar$\to ^{34}$Cl, and $^{46}$Cr$\to ^{46}$V.
I take the 1.8/2.0 (EM) interaction~\cite{Hebeler2011} with oscillator frequency $\hbar\omega=16$~MeV and $E_{\rm 3max}=16$.
The resulting $\delta_C$ values are plotted in Fig.~\ref{fig:deltaC3panel} as a function of the $e_{max}$ truncation.
I show results where the normal-ordering reference $|\Phi\rangle$ is taking to be either the inital or the final nucleus, and also the results from including only the one-body part of the evolved operator.
For reference, I also indicate the $\delta_C$ values adopted by Towner and Hardy~\cite{Hardy2015}.

If the calculation were under control, we should observe the following:
convergence with respect to $e_{\rm max}$; independence of the choice of reference; and a relatively small correction from including induced two-body terms, indicating a converging hierarchy of the cluster expansion.
For the $^{46}$Cr$\to^{46}$V transition, we observe reference independence and a small two-body correction, but only a hint of convergence in $e_{\rm max}$.
For the lighter nuclei the situation is worse, especially for $^{14}$O$\to^{14}$N.
In all cases, it appears that the large $e_{\rm max}$ behavior will need to be incorporated in some manner, possibly by utilizing natural orbitals~\cite{Tichai2019,Hoppe2020}, or by obtaining an extrapolation formula~\cite{Furnstahl2012,Furnstahl2015}.
However, I leave this for future work.
It appears that, at least for the near term, such calculations will tell us more about the IMSRG than about physics beyond the Standard Model.

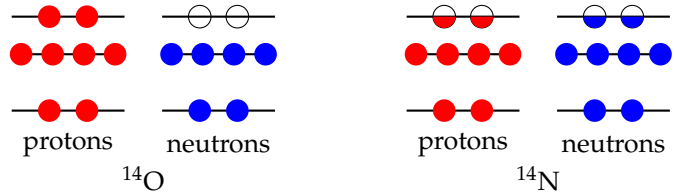
\begin{figure}
    \centering
    \begin{tikzpicture}[x=5mm,y=5mm]

\foreach \y in {0, 1.5,2.5 }{
 \draw [thick,black] (-3.5,\y) -- (-0.5,\y);
 \draw [thick,black] (0.5,\y) -- (3.5,\y);
};

\foreach \x in {-2.5,-1.5}{
  \fill [red] (\x,0) circle (1.5mm);
};
\foreach \x in {-3.25,-2.41,-1.58,-0.75}{
  \fill [red] (\x,1.5) circle (1.5mm);
};
\foreach \x in {-2.5,-1.5}{
  \fill [red] (\x,2.5) circle (1.5mm);
};

\foreach \x in {2.5,1.5}{
  \fill [blue] (\x,0) circle (1.5mm);
};
\foreach \x in {3.25,2.41,1.58,0.75}{
  \fill [blue] (\x,1.5) circle (1.5mm);
};

\foreach \x in {1.5,2.5}{
  \draw [black] (\x,2.5) circle (1.5mm);
  };

\node (o14) at (0,-1.8) {$^{14}$O};
\node (protons) at (-2,-0.9) {protons};
\node (neutrons) at (2,-0.9) {neutrons};

\end{tikzpicture}
    \hspace{4em}
    \begin{tikzpicture}[x=5mm,y=5mm]

\foreach \y in {0, 1.5,2.5 }{
 \draw [thick,black] (-3.5,\y) -- (-0.5,\y);
 \draw [thick,black] (0.5,\y) -- (3.5,\y);
};

\foreach \x in {-2.5,-1.5}{
  \fill [red] (\x,0) circle (1.5mm);
};
\foreach \x in {-3.25,-2.41,-1.58,-0.75}{
  \fill [red] (\x,1.5) circle (1.5mm);
};
\foreach \x in {-2.5,-1.5,1.5,2.5}{
  \draw [black] (\x,2.5) circle (1.5mm);
};

\foreach \x in {2.5,1.5}{
  \fill [blue] (\x,0) circle (1.5mm);
};
\foreach \x in {3.25,2.41,1.58,0.75}{
  \fill [blue] (\x,1.5) circle (1.5mm);
};
\foreach \x in {1.2,2.2}{
  \fill [blue] (\x,2.5) arc   (180:360:1.5mm) -- cycle;
};
\foreach \x in {-1.8,-2.8}{
  \fill [red] (\x,2.5) arc   (180:360:1.5mm) -- cycle;
};

\node (n14) at (0,-1.8) {$^{14}$N};
\node (protons) at (-2,-0.9) {protons};
\node (neutrons) at (2,-0.9) {neutrons};

\end{tikzpicture}
    \caption{Schematic illustration of reference states used for the $^{14}$O$\to^{14}$N decay in the left panel of Fig.~\ref{fig:deltaC3panel}. The half-filled circles for $^{14}$N indicate the equal filling of $m$-states used in the ensemble reference.}
    \label{fig:ShellsDiagram}
\end{figure}

The transition $^{14}$O$\to^{14}$N warrants a closer inspection, because it is light enough that it can be benchmarked against the no core shell model~\cite{Barrett2013}, although the observed $e_{\rm max}$ dependence suggests converged results may be challenging.
Moreover, the dramatic reference-dependence and contribution of two-body terms make this a good system for studying such effects, which are also important\footnote{In double beta decay, the bare operator is already two-body, and we are concerned with induced three-body terms, making this a more challenging case.} in neutrinoless double beta decay~\cite{Engel2017,Yao2020,Belley2020,Yao2021}.
The two references used are shown schematically in Fig.~\ref{fig:ShellsDiagram}.

Because $^{14}$O is a closed shell, it may also be treated by the coupled cluster method~\cite{Hagen2014} with the transition handled by the isospin-breaking equation-of-motion approach~\cite{Ekstrom2014}.
The strong $e_{\rm max}$ dependence suggests sensitivity to infrared physics, and this may be probed by including the continuum in the coupled cluster calculation~\cite{Hagen2010b}.
The results at $e_{\rm max}=12$ are indicated with yellow diamonds in Fig.~\ref{fig:deltaC3panel}.
The large  $\delta_C$ correction obtained with coupled cluster, as well as the substantial effect of including the continuum reinforces the notion that the pathology is not unique to the IMSRG.

The reference dependence persists even down to $e_{\rm max}=3$, which is sufficiently small that we may directly perform truncated configuration interaction (CI) calculations and extrapolate to the full CI result.
At $e_{\rm max}=3$, the VS-IMSRG yields $\delta_C$ values of 0.839\% and 0.056\% with $^{14}$O and $^{14}$N references, respectively.
The full CI result is 0.081\%, which is considerably closer to the $^{14}$N reference value.
This reinforces our suspicion that the calculation with the $^{14}$O reference is misbehaving.

As I mentioned above, results would be independent of the reference if all induced many-body terms were retained during the IMSRG evolution.
In practice we only retain up to two-body operators, so any reference dependence indicates the impact of discarded three-body or (if we are especially unlucky) higher-body terms.
If I truncate the expansion \eqref{eq:MagnusBCH} for the transformed operator at two nested commutators, the reference dependence remains, while truncating at one nested commutator eliminates the effect.
This suggests that three-body operators--- which first show up at two nested commutators if $\mathcal{O}(0)$ is purely one-body---are the culprit.

A more complete investigation would benefit from calculations at the IMSRG(3) level, which are becoming available~\cite{Heinz2021}, but which will not be pursued here.
For the moment, I will speculate.
It appears that using the $^{14}$O reference produces the largest error on the $\delta_C$ value out of all the calculations presented in Fig.~\ref{fig:deltaC3panel}.
This is naively counterintuitive because $^{14}$O is a closed shell and should be best approximated by the single-configuration reference.
But we are interested in the extent to which isospin symmetry is violated.
Choosing a reference that approximates $^{14}$O well but does a poor job for $^{14}$N artificially breaks this symmetry, leading to an overestimate of the correction $\delta_C$.
On the other hand, the ensemble $^{14}$N reference used is not a great approximation of the state of interest in $^{14}$N, which is actually open shell.
Using the $^{14}$N ensemble reference, we make an error in our description of $^{14}$N, and we make a similar error in the description of $^{14}$O, so the artificial breaking of isospin symmetry is reduced.
In support of this, using a $^{12}$C reference, which is a poor approximation of both $^{14}$O and $^{14}$N, results in $\delta_C$ values in very close agreement to those obtained with the $^{14}$N reference.

This is in line with the behavior observed with other relative quantities, namely excitation energies and separation energies.
It is a robust finding that the VS-IMSRG predicts too-high $2^+$ excitation energies for closed-shell nuclei~\cite{Simonis2017,Taniuchi2019}.
This can be understood by considering that the reference is a good approximation of the $0^+$ ground state, and a worse approximation for the $2^+$ excited state.
The truncation of three-body terms has a more severe impact on the $2^+$ state, leading to missed correlation energy, and consequently an excitation energy which is too high.
For open-shell nuclei, the reference is a mediocre approximation of both the ground and excited states, and so they are treated on more equal footing, leading to a more accurate excitation energy.
Likewise, it was observed that separation energies are more accurately obtained when using the same valence space for both the mass $A$ and mass $A-1$ nuclei, even if a different valence space might produce a more accurate ground state for one of the nuclei~\cite{Stroberg2021}.

Preliminary studies have found that including three-body terms ameliorates the issue with $2^+$ states, and an analogous effect is observed in coupled cluster~\cite{Hagen2016}.
Presumably retaining three-body terms would also reduce the dependence of separation energies on the choice of valence space, but this has not yet been explored.
One hopes that then including three-body terms will also help with the $\delta_C$ calculation.

\section{Conclusion and outlook}
We are in an exciting time in nuclear structure theory, in which it is becoming possible to address issues which were long plagued by ambiguities arising from inconsistent modelling.
On the question of Gamow-Teller decays, considerable progress has been made and it appears the issue is understood.
However, in order to fully put the question to rest one should properly assess theoretical errors from the EFT truncation and the many-body solution, and demonstrate systematic improvement in both.

In addition, a similar quenching is observed in strong-interaction charge exchange reactions~\cite{Gaarde1983}, where it is assumed that the transition operator for the target nucleus is proportional to $\sigma\tau$.
The quenching is often ascribed to missing strength at higher energies (the equivalent of ``correlations'' in beta decay).
However, three-body forces should play an analogous role as two-body currents do for beta decay (replace the axial current in the diagrams in Figure~\ref{fig:axialDiagrams} with an additional nucleon).
It will be interesting to quantitatively investigate this parallel.

For superallowed Fermi decays, the work of Towner and Hardy~\cite{Hardy2015} has laid a clear path, but there is more work to do on the many-body side.
A more detailed understanding of how errors creep into the calculations will be essential.

\funding{This work was supported by the U.S. Department of Energy, Office of Science, Office of Nuclear Physics, contract no. DEAC02-06CH11357}
%

%
\acknowledgments{I thank my collaborators P.~Gysbers, G.~Hagen, J.~D.~Holt, G.~R.~Jansen, T.~D.~Morris, P.~Navr\'atil, T.~Papenbrock, S.~Quaglioni, A.~Schwenk, and K.~A.~Wendt for providing matrix elements of the chiral NN+3N interactions and axial vector currents.
I also thank T.~Miyagi for helpful discussions regarding the KSHELL code and various related scripts.}
%
%
%
%
%
%
\appendixstart
\appendix
\section{Table of Gamow-Teller matrix elements}

\begin{specialtable}[]
    \caption{Matrix elements for the Gamow-Teller transitions plotted in Figures~\ref{fig:GT2panel} and ~\ref{fig:GTq_vs_A}. In the last column, $\dagger$ indicates the lowest 3 states of the listed $J^{\pi}$ in the final nucleus are summed in the matrix element, while $*$ indicates that 4 states are summed.}
    \label{tab:GTresults}
    \centering
    \begin{tabular}{ccccccccl}
\hline \hline \\
$A$ & $Z_i$ & $Z_f$ & $2J_i$ & $2J_f$ & $M(GT)_{exp}$ & $\sigma\tau_{bare}$ & $\sigma\tau_{IMSRG}$ & $M(GT)_{th}$  \\
\hline \\
6 & 2 & 3 & 0 & 2 & 2.748  &  2.995 & 2.817  & 2.776   \\
7 & 4 & 3 & 3 & 3 & 2.882  &  3.088 & 2.889  & 2.808   \\
7 & 4 & 3 & 3 & 1 & 2.678  &  2.907 & 2.709  & 2.639   \\
12 & 7 & 6 & 2 & 0 & 1.184  &  0.830 & 0.637  & 0.616   \\
12 & 7 & 6 & 2 & 2 & 2.370  &  1.816 & 1.728  & 1.687   \\
15 & 8 & 7 & 1 & 1 & 0.889  &  1.037 & 1.077  & 1.035   \\
17 & 9 & 8 & 5 & 5 & 3.168  &  3.681 & 3.504  & 3.424   \\
18 & 9 & 8 & 2 & 0 & 2.209  &  2.860 & 2.483  & 2.430   \\
19 & 10 & 9 & 1 & 1 & 2.273  &  2.944 & 2.641  & 2.540   \\
20 & 8 & 9 & 0 & 2 & 1.058  &  1.419 & 1.307  & 1.266   \\
20 & 11 & 10 & 4 & 2 & 2.403  &  2.430 & 2.208  & 2.124   \\
24 & 13 & 12 & 8 & 8 & 2.886  &  3.023 & 2.731  & 2.639 $\dagger$ \\
25 & 13 & 12 & 5 & 5 & 1.971  &  2.433 & 2.273  & 2.139   \\
26 & 14 & 13 & 0 & 2 & 3.055  &  3.648 & 3.128  & 3.006  $*$ \\ 
27 & 11 & 12 & 5 & 3 & 1.361  &  1.718 & 1.435  & 1.365   \\
27 & 14 & 13 & 5 & 5 & 1.688  &  1.934 & 1.703  & 1.630   \\
30 & 12 & 13 & 0 & 2 & 1.090  &  1.488 & 1.281  & 1.181   \\
35 & 15 & 16 & 1 & 1 & 1.033  &  1.299 & 1.060  & 0.999   \\
37 & 19 & 18 & 3 & 3 & 1.169  &  1.768 & 1.633  & 1.565   \\
39 & 20 & 19 & 3 & 3 & 1.308  &  1.967 & 1.724  & 1.628   \\
41 & 21 & 20 & 7 & 7 & 2.999  &  4.073 & 3.733  & 3.548   \\
42 & 21 & 20 & 14 & 12 & 2.497  &  3.305 & 3.006  & 2.836   \\
42 & 22 & 21 & 0 & 2 & 2.038  &  2.713 & 2.415  & 2.237   \\
45 & 22 & 21 & 7 & 7 & 1.123  &  1.513 & 1.362  & 1.252   \\
45 & 23 & 22 & 7 & 7 & 1.801  &  2.177 & 1.982  & 1.844   \\
47 & 24 & 23 & 3 & 3 & 0.942  &  1.190 & 1.077  & 0.967   \\
48 & 25 & 24 & 8 & 8 & 3.596  &  4.081 & 3.494  & 3.340 $*$ \\ 
49 & 25 & 24 & 5 & 5 & 1.364  &  1.768 & 1.525  & 1.466   \\
49 & 25 & 24 & 5 & 7 & 0.764  &  0.768 & 0.656  & 0.629   \\
100 & 50 & 49 & 0 & 2 & 2.870  &  5.355 & 3.717  & 3.471   \\
\hline \hline  \\
\end{tabular}
\end{specialtable}

%
%
%
\end{paracol}
\reftitle{References}


\externalbibliography{yes}
\bibliography{references.bib}

\begin{thebibliography}{999}

\bibitem[Fermi(1934)]{Fermi1934}
Fermi, E.
\newblock {Versuch einer Theorie der $\beta$-Strahlen. I}.
\newblock {\em Zeitschrift f\"ur Phys.} {\bf 1934}, {\em 88},~161--177.
\newblock
  doi:{\changeurlcolor{black}\href{https://doi.org/10.1007/BF01351864}{\detokenize{10.1007/BF01351864}}}.

\bibitem[Lee and Yang(1956)]{Lee1956}
Lee, T.D.; Yang, C.N.
\newblock {Question of Parity Conservation in Weak Interactions}.
\newblock {\em Phys. Rev.} {\bf 1956}, {\em 104},~254--258.

\bibitem[Wu \em{et~al.}(1957)Wu, Ambler, Hayward, Hoppes, and Hudson]{Wu1957}
Wu, C.S.; Ambler, E.; Hayward, R.W.; Hoppes, D.D.; Hudson, R.P.
\newblock {Experimental test of parity conservation in beta decay [5]}.
\newblock {\em Phys. Rev.} {\bf 1957}, {\em 105},~1413--1415.
\newblock
  doi:{\changeurlcolor{black}\href{https://doi.org/10.1103/PhysRev.105.1413}{\detokenize{10.1103/PhysRev.105.1413}}}.

\bibitem[Avignone \em{et~al.}(2008)Avignone, Elliott, and Engel]{Avignone2008}
Avignone, F.T.; Elliott, S.R.; Engel, J.
\newblock {Double beta decay, Majorana neutrinos, and neutrino mass}.
\newblock {\em Rev. Mod. Phys.} {\bf 2008}, {\em 80},~481--516.
\newblock
  doi:{\changeurlcolor{black}\href{https://doi.org/10.1103/RevModPhys.80.481}{\detokenize{10.1103/RevModPhys.80.481}}}.

\bibitem[Holstein(2014)]{Holstein2014}
Holstein, B.R.
\newblock {Precision frontier in semileptonic weak interactions: Theory}.
\newblock {\em J. Phys. G Nucl. Part. Phys.} {\bf 2014}, {\em 41}.
\newblock
  doi:{\changeurlcolor{black}\href{https://doi.org/10.1088/0954-3899/41/11/114001}{\detokenize{10.1088/0954-3899/41/11/114001}}}.

\bibitem[Hardy and Towner(2015)]{Hardy2015}
Hardy, J.C.; Towner, I.S.
\newblock {Superallowed $0^+\to0^+$ nuclear $\beta$ decays: 2014 critical
  survey, with precise results for $V_{ud}$ and CKM unitarity}.
\newblock {\em Phys. Rev. C} {\bf 2015}, {\em 91},~025501.
\newblock
  doi:{\changeurlcolor{black}\href{https://doi.org/10.1103/PhysRevC.91.025501}{\detokenize{10.1103/PhysRevC.91.025501}}}.

\bibitem[Hayen \em{et~al.}(2018)Hayen, Severijns, Bodek, Rozpedzik, and
  Mougeot]{Hayen2018}
Hayen, L.; Severijns, N.; Bodek, K.; Rozpedzik, D.; Mougeot, X.
\newblock {High precision analytical description of the allowed $\beta$
  spectrum shape}.
\newblock {\em Rev. Mod. Phys.} {\bf 2018}, {\em 90},~015008.
\newblock
  doi:{\changeurlcolor{black}\href{https://doi.org/10.1103/RevModPhys.90.015008}{\detokenize{10.1103/RevModPhys.90.015008}}}.

\bibitem[Epelbaum \em{et~al.}(2009)Epelbaum, Hammer, and
  Mei{\ss}ner]{Epelbaum2009}
Epelbaum, E.; Hammer, H.W.; Mei{\ss}ner, U.G.
\newblock {Modern theory of nuclear forces}.
\newblock {\em Rev. Mod. Phys.} {\bf 2009}, {\em 81},~1773--1825.
\newblock
  doi:{\changeurlcolor{black}\href{https://doi.org/10.1103/RevModPhys.81.1773}{\detokenize{10.1103/RevModPhys.81.1773}}}.

\bibitem[Machleidt and Entem(2011)]{Machleidt2011}
Machleidt, R.; Entem, D.R.
\newblock {Chiral effective field theory and nuclear forces}.
\newblock {\em Phys. Rep.} {\bf 2011}, {\em 503},~1--75.
\newblock
  doi:{\changeurlcolor{black}\href{https://doi.org/10.1016/j.physrep.2011.02.001}{\detokenize{10.1016/j.physrep.2011.02.001}}}.

\bibitem[Hammer \em{et~al.}(2020)Hammer, K{\"{o}}nig, and van
  Kolck]{Hammer2019}
Hammer, H.W.; K{\"{o}}nig, S.; van Kolck, U.
\newblock {Nuclear effective field theory: Status and perspectives}.
\newblock {\em Rev. Mod. Phys.} {\bf 2020}, {\em 92},~025004,
  \href{http://xxx.lanl.gov/abs/1906.12122}{{\normalfont [1906.12122]}}.
\newblock
  doi:{\changeurlcolor{black}\href{https://doi.org/10.1103/RevModPhys.92.025004}{\detokenize{10.1103/RevModPhys.92.025004}}}.

\bibitem[Carlson \em{et~al.}(2015)Carlson, Gandolfi, Pederiva, Pieper,
  Schiavilla, Schmidt, and Wiringa]{Carlson2015}
Carlson, J.; Gandolfi, S.; Pederiva, F.; Pieper, S.C.; Schiavilla, R.; Schmidt,
  K.E.E.; Wiringa, R.B.B.
\newblock {Quantum Monte Carlo methods for nuclear physics}.
\newblock {\em Rev. Mod. Phys.} {\bf 2015}, {\em 87},~1067--1118.
\newblock
  doi:{\changeurlcolor{black}\href{https://doi.org/10.1103/RevModPhys.87.1067}{\detokenize{10.1103/RevModPhys.87.1067}}}.

\bibitem[Hagen \em{et~al.}(2014)Hagen, Papenbrock, Hjorth-Jensen, and
  Dean]{Hagen2014}
Hagen, G.; Papenbrock, T.; Hjorth-Jensen, M.; Dean, D.J.
\newblock {Coupled-cluster computations of atomic nuclei}.
\newblock {\em Reports Prog. Phys.} {\bf 2014}, {\em 77},~096302.
\newblock
  doi:{\changeurlcolor{black}\href{https://doi.org/10.1088/0034-4885/77/9/096302}{\detokenize{10.1088/0034-4885/77/9/096302}}}.

\bibitem[Binder \em{et~al.}(2014)Binder, Langhammer, Calci, and
  Roth]{Binder2014}
Binder, S.; Langhammer, J.; Calci, A.; Roth, R.
\newblock {Ab initio path to heavy nuclei}.
\newblock {\em Phys. Lett. B} {\bf 2014}, {\em 736},~119.
\newblock
  doi:{\changeurlcolor{black}\href{https://doi.org/10.1016/j.physletb.2014.07.010}{\detokenize{10.1016/j.physletb.2014.07.010}}}.

\bibitem[Morris \em{et~al.}(2018)Morris, Simonis, Stroberg, Stumpf, Hagen,
  Holt, Jansen, Papenbrock, Roth, and Schwenk]{Morris2018}
Morris, T.D.; Simonis, J.; Stroberg, S.R.; Stumpf, C.; Hagen, G.; Holt, J.D.;
  Jansen, G.R.; Papenbrock, T.; Roth, R.; Schwenk, A.
\newblock {Structure of the Lightest Tin Isotopes}.
\newblock {\em Phys. Rev. Lett.} {\bf 2018}, {\em 120},~152503,
  \href{http://xxx.lanl.gov/abs/1709.02786}{{\normalfont [1709.02786]}}.
\newblock
  doi:{\changeurlcolor{black}\href{https://doi.org/10.1103/PhysRevLett.120.152503}{\detokenize{10.1103/PhysRevLett.120.152503}}}.

\bibitem[Hergert(2020)]{Hergert2020}
Hergert, H.
\newblock {A Guided Tour of ab initio Nuclear Many-Body Theory}.
\newblock {\em Front. Phys.} {\bf 2020}, {\em 8},~1--33,
  \href{http://xxx.lanl.gov/abs/2008.05061}{{\normalfont [2008.05061]}}.
\newblock
  doi:{\changeurlcolor{black}\href{https://doi.org/10.3389/fphy.2020.00379}{\detokenize{10.3389/fphy.2020.00379}}}.

\bibitem[Som{\`{a}}(2020)]{Soma2020a}
Som{\`{a}}, V.
\newblock {Self-Consistent Green's Function Theory for Atomic Nuclei}.
\newblock {\em Front. Phys.} {\bf 2020}, {\em 8},~1--31,
  \href{http://xxx.lanl.gov/abs/2003.11321}{{\normalfont [2003.11321]}}.
\newblock
  doi:{\changeurlcolor{black}\href{https://doi.org/10.3389/fphy.2020.00340}{\detokenize{10.3389/fphy.2020.00340}}}.

\bibitem[Lee(2020)]{Lee2020}
Lee, D.
\newblock {Recent Progress in Nuclear Lattice Simulations}.
\newblock {\em Front. Phys.} {\bf 2020}, {\em 8},~1--7.
\newblock
  doi:{\changeurlcolor{black}\href{https://doi.org/10.3389/fphy.2020.00174}{\detokenize{10.3389/fphy.2020.00174}}}.

\bibitem[van Kolck(2020)]{VanKolck2020}
van Kolck, U.
\newblock {The Problem of Renormalization of Chiral Nuclear Forces} {\bf 2020}.
\newblock  \href{http://xxx.lanl.gov/abs/2003.06721}{{\normalfont
  [2003.06721]}}.

\bibitem[Phillips(2021)]{Phillips2021}
Phillips, D.R.
\newblock {What hath Weinberg wrought? Reflections on what Weinberg's papers on
  'Nuclear Forces from Chiral Lagrangians' did and did not accomplish} {\bf
  2021}.
\newblock pp. 1--19,  \href{http://xxx.lanl.gov/abs/2107.03558}{{\normalfont
  [2107.03558]}}.

\bibitem[Cirigliano \em{et~al.}(2021)Cirigliano, Dekens, de~Vries, Hoferichter,
  and Mereghetti]{Cirigliano2021}
Cirigliano, V.; Dekens, W.; de~Vries, J.; Hoferichter, M.; Mereghetti, E.
\newblock {Toward Complete Leading-Order Predictions for Neutrinoless Double
  $\beta$ Decay}.
\newblock {\em Phys. Rev. Lett.} {\bf 2021}, {\em 126},~172002,
  \href{http://xxx.lanl.gov/abs/2012.11602}{{\normalfont [2012.11602]}}.
\newblock
  doi:{\changeurlcolor{black}\href{https://doi.org/10.1103/PhysRevLett.126.172002}{\detokenize{10.1103/PhysRevLett.126.172002}}}.

\bibitem[Bogner \em{et~al.}(2010)Bogner, Furnstahl, and Schwenk]{Bogner2010}
Bogner, S.; Furnstahl, R.; Schwenk, A.
\newblock {From low-momentum interactions to nuclear structure}.
\newblock {\em Prog. Part. Nucl. Phys.} {\bf 2010}, {\em 65},~94--147.
\newblock
  doi:{\changeurlcolor{black}\href{https://doi.org/10.1016/j.ppnp.2010.03.001}{\detokenize{10.1016/j.ppnp.2010.03.001}}}.

\bibitem[Furnstahl and Hebeler(2013)]{Furnstahl2013}
Furnstahl, R.J.; Hebeler, K.
\newblock {New applications of renormalization group methods in nuclear
  physics}.
\newblock {\em Reports Prog. Phys.} {\bf 2013}, {\em 76},~126301.
\newblock
  doi:{\changeurlcolor{black}\href{https://doi.org/10.1088/0034-4885/76/12/126301}{\detokenize{10.1088/0034-4885/76/12/126301}}}.

\bibitem[Hergert \em{et~al.}(2016)Hergert, Bogner, Morris, Schwenk, and
  Tsukiyama]{Hergert2016}
Hergert, H.; Bogner, S.K.; Morris, T.D.; Schwenk, A.; Tsukiyama, K.
\newblock {The In-Medium Similarity Renormalization Group: A Novel Ab Initio
  Method for Nuclei}.
\newblock {\em Phys. Rep.} {\bf 2016}, {\em 621},~165--222.
\newblock
  doi:{\changeurlcolor{black}\href{https://doi.org/10.1016/j.physrep.2015.12.007}{\detokenize{10.1016/j.physrep.2015.12.007}}}.

\bibitem[Hergert(2017)]{Hergert2017}
Hergert, H.
\newblock {In-Medium Similarity Renormalization Group for Closed and Open-Shell
  Nuclei}.
\newblock {\em Phys. Scr.} {\bf 2017}, {\em 92},~023002,
  \href{http://xxx.lanl.gov/abs/1607.06882}{{\normalfont [1607.06882]}}.
\newblock
  doi:{\changeurlcolor{black}\href{https://doi.org/10.1088/1402-4896/92/2/023002}{\detokenize{10.1088/1402-4896/92/2/023002}}}.

\bibitem[Hergert \em{et~al.}(2017)Hergert, Bogner, Lietz, Morris, Novario,
  Parzuchowski, and Yuan]{Hergert2017a}
Hergert, H.; Bogner, S.K.; Lietz, J.G.; Morris, T.D.; Novario, S.J.;
  Parzuchowski, N.M.; Yuan, F.
\newblock {In-Medium Similarity Renormalization Group Approach to the Nuclear
  Many-Body Problem}. In {\em Lect. Notes Phys.}; Hjorth-Jensen, M.; Lombardo,
  M.P.; van Kolck, U., Eds.; Springer,  2017; pp. 477--570,
  \href{http://xxx.lanl.gov/abs/1612.08315}{{\normalfont [1612.08315]}}.
\newblock
  doi:{\changeurlcolor{black}\href{https://doi.org/10.1007/978-3-319-53336-0_10}{\detokenize{10.1007/978-3-319-53336-0_10}}}.

\bibitem[Stroberg \em{et~al.}(2019)Stroberg, Bogner, Hergert, and
  Holt]{Stroberg2019}
Stroberg, S.R.; Bogner, S.K.; Hergert, H.; Holt, J.D.
\newblock {Nonempirical Interactions for the Nuclear Shell Model: An Update}.
\newblock {\em Annu. Rev. Nucl. Part. Sci.} {\bf 2019}, {\em 69},~307--362,
  \href{http://xxx.lanl.gov/abs/1902.06154}{{\normalfont [1902.06154]}}.
\newblock
  doi:{\changeurlcolor{black}\href{https://doi.org/10.1146/annurev-nucl-101917-021120}{\detokenize{10.1146/annurev-nucl-101917-021120}}}.

\bibitem[Anderson \em{et~al.}(2010)Anderson, Bogner, Furnstahl, and
  Perry]{Anderson2010}
Anderson, E.R.; Bogner, S.K.; Furnstahl, R.J.; Perry, R.J.
\newblock {Operator evolution via the similarity renormalization group: The
  deuteron}.
\newblock {\em Phys. Rev. C} {\bf 2010}, {\em 82},~054001.
\newblock
  doi:{\changeurlcolor{black}\href{https://doi.org/10.1103/PhysRevC.82.054001}{\detokenize{10.1103/PhysRevC.82.054001}}}.

\bibitem[Schuster \em{et~al.}(2014)Schuster, Quaglioni, Johnson, Jurgenson, and
  Navr{\'{a}}til]{Schuster2014}
Schuster, M.D.; Quaglioni, S.; Johnson, C.W.; Jurgenson, E.D.; Navr{\'{a}}til,
  P.
\newblock {Operator evolution for ab initio theory of light nuclei}.
\newblock {\em Phys. Rev. C} {\bf 2014}, {\em 90},~011301.
\newblock
  doi:{\changeurlcolor{black}\href{https://doi.org/10.1103/PhysRevC.90.011301}{\detokenize{10.1103/PhysRevC.90.011301}}}.

\bibitem[Parzuchowski \em{et~al.}(2017)Parzuchowski, Stroberg, Navr{\'{a}}til,
  Hergert, and Bogner]{Parzuchowski2017}
Parzuchowski, N.M.; Stroberg, S.R.; Navr{\'{a}}til, P.; Hergert, H.; Bogner,
  S.K.
\newblock {Ab initio electromagnetic observables with the in-medium similarity
  renormalization group}.
\newblock {\em Phys. Rev. C} {\bf 2017}, {\em 96},~034324,
  \href{http://xxx.lanl.gov/abs/1705.05511}{{\normalfont [1705.05511]}}.
\newblock
  doi:{\changeurlcolor{black}\href{https://doi.org/10.1103/PhysRevC.96.034324}{\detokenize{10.1103/PhysRevC.96.034324}}}.

\bibitem[Tropiano \em{et~al.}(2020)Tropiano, Bogner, and
  Furnstahl]{Tropiano2020}
Tropiano, A.J.; Bogner, S.K.; Furnstahl, R.J.
\newblock {Operator evolution from the similarity renormalization group and the
  Magnus expansion}.
\newblock {\em Phys. Rev. C} {\bf 2020}, {\em 102},~034005,
  \href{http://xxx.lanl.gov/abs/2006.11186}{{\normalfont [2006.11186]}}.
\newblock
  doi:{\changeurlcolor{black}\href{https://doi.org/10.1103/PhysRevC.102.034005}{\detokenize{10.1103/PhysRevC.102.034005}}}.

\bibitem[White(2002)]{White2002}
White, S.R.
\newblock {Numerical canonical transformation approach to quantum many-body
  problems}.
\newblock {\em J. Chem. Phys.} {\bf 2002}, {\em 117},~7472.
\newblock
  doi:{\changeurlcolor{black}\href{https://doi.org/10.1063/1.1508370}{\detokenize{10.1063/1.1508370}}}.

\bibitem[Stroberg \em{et~al.}(2017)Stroberg, Calci, Hergert, Holt, Bogner,
  Roth, and Schwenk]{Stroberg2017}
Stroberg, S.R.; Calci, A.; Hergert, H.; Holt, J.D.; Bogner, S.K.; Roth, R.;
  Schwenk, A.
\newblock {Nucleus-Dependent Valence-Space Approach to Nuclear Structure}.
\newblock {\em Phys. Rev. Lett.} {\bf 2017}, {\em 118},~032502,
  \href{http://xxx.lanl.gov/abs/1607.03229}{{\normalfont [1607.03229]}}.
\newblock
  doi:{\changeurlcolor{black}\href{https://doi.org/10.1103/PhysRevLett.118.032502}{\detokenize{10.1103/PhysRevLett.118.032502}}}.

\bibitem[Morris \em{et~al.}(2015)Morris, Parzuchowski, and Bogner]{Morris2015}
Morris, T.D.; Parzuchowski, N.M.; Bogner, S.K.
\newblock {Magnus expansion and in-medium similarity renormalization group}.
\newblock {\em Phys. Rev. C} {\bf 2015}, {\em 92},~034331.
\newblock
  doi:{\changeurlcolor{black}\href{https://doi.org/10.1103/PhysRevC.92.034331}{\detokenize{10.1103/PhysRevC.92.034331}}}.

\bibitem[Wilkinson(1973{\natexlab{a}})]{Wilkinson1973a}
Wilkinson, D.H.
\newblock {Renormalization of the Axial-Vector Coupling Constant in Nuclear
  $\beta$ Decay}.
\newblock {\em Phys. Rev. C} {\bf 1973}, {\em 7},~930--936.
\newblock
  doi:{\changeurlcolor{black}\href{https://doi.org/10.1103/PhysRevC.7.930}{\detokenize{10.1103/PhysRevC.7.930}}}.

\bibitem[Wilkinson(1973{\natexlab{b}})]{Wilkinson1973}
Wilkinson, D.
\newblock {Renormalization of the axial-vector coupling constant in nuclear
  $\beta$-decay (II)}.
\newblock {\em Nucl. Phys. A} {\bf 1973}, {\em 209},~470--484.
\newblock
  doi:{\changeurlcolor{black}\href{https://doi.org/10.1016/0375-9474(73)90840-3}{\detokenize{10.1016/0375-9474(73)90840-3}}}.

\bibitem[Brown \em{et~al.}(1978)Brown, Chung, and Wildenthal]{Brown1978}
Brown, B.A.; Chung, W.; Wildenthal, B.H.
\newblock {Empirical Renormalization of the One-Body Gamow-Teller $\beta$
  -Decay Matrix Elements in the 1s-0d Shell}.
\newblock {\em Phys. Rev. Lett.} {\bf 1978}, {\em 40},~1631--1635.
\newblock
  doi:{\changeurlcolor{black}\href{https://doi.org/10.1103/PhysRevLett.40.1631}{\detokenize{10.1103/PhysRevLett.40.1631}}}.

\bibitem[Brown and Wildenthal(1985)]{Brown1985a}
Brown, B.; Wildenthal, B.
\newblock {Experimental and theoretical Gamow-Teller beta-decay observables for
  the sd-shell nuclei}.
\newblock {\em At. Data Nucl. Data Tables} {\bf 1985}, {\em 33},~347--404.
\newblock
  doi:{\changeurlcolor{black}\href{https://doi.org/10.1016/0092-640X(85)90009-9}{\detokenize{10.1016/0092-640X(85)90009-9}}}.

\bibitem[Chou \em{et~al.}(1993)Chou, Warburton, and Brown]{Chou1993}
Chou, W.T.; Warburton, E.K.; Brown, B.A.
\newblock {Gamow-Teller beta-decay rates for A$\leq$18 nuclei}.
\newblock {\em Phys. Rev. C} {\bf 1993}, {\em 47},~163--177.
\newblock
  doi:{\changeurlcolor{black}\href{https://doi.org/10.1103/PhysRevC.47.163}{\detokenize{10.1103/PhysRevC.47.163}}}.

\bibitem[Mart{\'{i}}nez-Pinedo \em{et~al.}(1996)Mart{\'{i}}nez-Pinedo, Poves,
  Caurier, and Zuker]{Martinez-Pinedo1996}
Mart{\'{i}}nez-Pinedo, G.; Poves, A.; Caurier, E.; Zuker, A.P.
\newblock {Effective $g_A$ in the $pf$ shell}.
\newblock {\em Phys. Rev. C} {\bf 1996}, {\em 53},~R2602.
\newblock
  doi:{\changeurlcolor{black}\href{https://doi.org/10.1103/PhysRevC.53.R2602}{\detokenize{10.1103/PhysRevC.53.R2602}}}.

\bibitem[Rho(1974)]{Rho1974}
Rho, M.
\newblock {Quenching of axial-vector coupling constant in $\beta$-decay and
  pion-nucleus optical potential}.
\newblock {\em Nucl. Physics, Sect. A} {\bf 1974}, {\em 231},~493--503.
\newblock
  doi:{\changeurlcolor{black}\href{https://doi.org/10.1016/0375-9474(74)90512-0}{\detokenize{10.1016/0375-9474(74)90512-0}}}.

\bibitem[Towner and Khanna(1983)]{Towner1983}
Towner, I.S.; Khanna, F.C.
\newblock {Corrections to the single-particle M1 and Gamow-Teller matrix
  elements}.
\newblock {\em Nucl. Phys. A} {\bf 1983}, {\em 399},~334--364.

\bibitem[Ericson \em{et~al.}(1973)Ericson, Figureau, and
  Th{\'{e}}venet]{Ericson1973}
Ericson, M.; Figureau, A.; Th{\'{e}}venet, C.
\newblock {Pionic field and renormalization of the axial coupling constant in
  nuclei}.
\newblock {\em Phys. Lett. B} {\bf 1973}, {\em 45},~19--22.
\newblock
  doi:{\changeurlcolor{black}\href{https://doi.org/10.1016/0370-2693(73)90242-6}{\detokenize{10.1016/0370-2693(73)90242-6}}}.

\bibitem[Scherer and Schindler(2012)]{Scherer2012}
Scherer, S.; Schindler, M.R.
\newblock {\em {A Primer for Chiral Perturbation Theory}}; Vol. 830, {\em
  Lecture Notes in Physics}, Springer Berlin Heidelberg: Berlin, Heidelberg,
  2012.
\newblock
  doi:{\changeurlcolor{black}\href{https://doi.org/10.1007/978-3-642-19254-8}{\detokenize{10.1007/978-3-642-19254-8}}}.

\bibitem[Park \em{et~al.}(2003)Park, Marcucci, Schiavilla, Viviani, Kievsky,
  Rosati, Kubodera, Min, and Rho]{Park2003}
Park, T.S.; Marcucci, L.E.; Schiavilla, R.; Viviani, M.; Kievsky, A.; Rosati,
  S.; Kubodera, K.; Min, D.P.; Rho, M.
\newblock {Parameter-free effective field theory calculation for the solar
  proton-fusion and hep processes}.
\newblock {\em Phys. Rev. C} {\bf 2003}, {\em 67},~055206.
\newblock
  doi:{\changeurlcolor{black}\href{https://doi.org/10.1103/PhysRevC.67.055206}{\detokenize{10.1103/PhysRevC.67.055206}}}.

\bibitem[G{\aa}rdestig and Phillips(2006)]{Gardestig2006}
G{\aa}rdestig, A.; Phillips, D.R.
\newblock {How Low-Energy Weak Reactions Can Constrain Three-Nucleon Forces and
  the Neutron-Neutron Scattering Length}.
\newblock {\em Phys. Rev. Lett.} {\bf 2006}, {\em 96},~232301.
\newblock
  doi:{\changeurlcolor{black}\href{https://doi.org/10.1103/PhysRevLett.96.232301}{\detokenize{10.1103/PhysRevLett.96.232301}}}.

\bibitem[Gazit \em{et~al.}(2009)Gazit, Quaglioni, and
  Navr{\'{a}}til]{Gazit2009}
Gazit, D.; Quaglioni, S.; Navr{\'{a}}til, P.
\newblock {Three-Nucleon Low-Energy Constants from the Consistency of
  Interactions and Currents in Chiral Effective Field Theory}.
\newblock {\em Phys. Rev. Lett.} {\bf 2009}, {\em 103},~102502.
\newblock
  doi:{\changeurlcolor{black}\href{https://doi.org/10.1103/PhysRevLett.103.102502}{\detokenize{10.1103/PhysRevLett.103.102502}}}.

\bibitem[Krebs \em{et~al.}(2017)Krebs, Epelbaum, and Mei{\ss}ner]{Krebs2017}
Krebs, H.; Epelbaum, E.; Mei{\ss}ner, U.G.G.
\newblock {Nuclear axial current operators to fourth order in chiral effective
  field theory}.
\newblock {\em Ann. Phys. (N. Y).} {\bf 2017}, {\em 378},~317--395,
  \href{http://xxx.lanl.gov/abs/1610.03569}{{\normalfont [1610.03569]}}.
\newblock
  doi:{\changeurlcolor{black}\href{https://doi.org/10.1016/j.aop.2017.01.021}{\detokenize{10.1016/j.aop.2017.01.021}}}.

\bibitem[Park \em{et~al.}(1997)Park, Jung, and Min]{Park1997}
Park, T.S.; Jung, H.; Min, D.P.
\newblock {In-medium effective axial-vector coupling constant}.
\newblock {\em Phys. Lett. Sect. B Nucl. Elem. Part. High-Energy Phys.} {\bf
  1997}, {\em 409},~26--32.
\newblock
  doi:{\changeurlcolor{black}\href{https://doi.org/10.1016/S0370-2693(97)00880-0}{\detokenize{10.1016/S0370-2693(97)00880-0}}}.

\bibitem[Men{\'{e}}ndez \em{et~al.}(2011)Men{\'{e}}ndez, Gazit, and
  Schwenk]{Menendez2011}
Men{\'{e}}ndez, J.; Gazit, D.; Schwenk, A.
\newblock {Chiral Two-Body Currents in Nuclei: Gamow-Teller Transitions and
  Neutrinoless Double-Beta Decay}.
\newblock {\em Phys. Rev. Lett.} {\bf 2011}, {\em 107},~062501.
\newblock
  doi:{\changeurlcolor{black}\href{https://doi.org/10.1103/PhysRevLett.107.062501}{\detokenize{10.1103/PhysRevLett.107.062501}}}.

\bibitem[Gysbers \em{et~al.}(2019)Gysbers, Hagen, Holt, Jansen, Morris,
  Navr{\'{a}}til, Papenbrock, Quaglioni, Schwenk, Stroberg, and
  Wendt]{Gysbers2019}
Gysbers, P.; Hagen, G.; Holt, J.D.; Jansen, G.R.; Morris, T.D.; Navr{\'{a}}til,
  P.; Papenbrock, T.; Quaglioni, S.; Schwenk, A.; Stroberg, S.R.; Wendt, K.A.
\newblock {Discrepancy between experimental and theoretical $\beta$-decay rates
  resolved from first principles}.
\newblock {\em Nat. Phys.} {\bf 2019}, {\em 15},~428--431,
  \href{http://xxx.lanl.gov/abs/arXiv:1903.00047v1}{{\normalfont
  [arXiv:1903.00047v1]}}.
\newblock
  doi:{\changeurlcolor{black}\href{https://doi.org/10.1038/s41567-019-0450-7}{\detokenize{10.1038/s41567-019-0450-7}}}.

\bibitem[Ekstr{\"{o}}m \em{et~al.}(2014)Ekstr{\"{o}}m, Jansen, Wendt, Hagen,
  Papenbrock, Bacca, Carlsson, and Gazit]{Ekstrom2014}
Ekstr{\"{o}}m, A.; Jansen, G.R.; Wendt, K.A.; Hagen, G.; Papenbrock, T.; Bacca,
  S.; Carlsson, B.; Gazit, D.
\newblock {Effects of three-nucleon forces and two-body currents on
  Gamow-Teller strengths.}
\newblock {\em Phys. Rev. Lett.} {\bf 2014}, {\em 113},~262504.
\newblock
  doi:{\changeurlcolor{black}\href{https://doi.org/10.1103/PhysRevLett.113.262504}{\detokenize{10.1103/PhysRevLett.113.262504}}}.

\bibitem[Gazit \em{et~al.}(2019)Gazit, Quaglioni, and
  Navr{\'{a}}til]{Gazit2019}
Gazit, D.; Quaglioni, S.; Navr{\'{a}}til, P.
\newblock {Erratum: Three-Nucleon Low-Energy Constants from the Consistency of
  Interactions and Currents in Chiral Effective Field Theory [Phys. Rev. Lett.
  103 , 102502 (2009)]}.
\newblock {\em Phys. Rev. Lett.} {\bf 2019}, {\em 122},~029901.
\newblock
  doi:{\changeurlcolor{black}\href{https://doi.org/10.1103/PhysRevLett.122.029901}{\detokenize{10.1103/PhysRevLett.122.029901}}}.

\bibitem[Holt \em{et~al.}(2008)Holt, Brown, Kuo, Holt, and Machleidt]{Holt2008}
Holt, J.W.; Brown, G.E.; Kuo, T.T.S.; Holt, J.D.; Machleidt, R.
\newblock {Shell model description of the $^{14}$C dating $\beta$ decay with
  Brown-Rho-scaled NN interactions}.
\newblock {\em Phys. Rev. Lett.} {\bf 2008}, {\em 100},~1--4,
  \href{http://xxx.lanl.gov/abs/arXiv:0710.0310v2}{{\normalfont
  [arXiv:0710.0310v2]}}.
\newblock
  doi:{\changeurlcolor{black}\href{https://doi.org/10.1103/PhysRevLett.100.062501}{\detokenize{10.1103/PhysRevLett.100.062501}}}.

\bibitem[Maris \em{et~al.}(2011)Maris, Vary, Navr{\'{a}}til, Ormand, Nam, and
  Dean]{Maris2011}
Maris, P.; Vary, J.P.; Navr{\'{a}}til, P.; Ormand, W.E.; Nam, H.; Dean, D.J.
\newblock {Origin of the Anomalous Long Lifetime of C 14}.
\newblock {\em Phys. Rev. Lett.} {\bf 2011}, {\em 106},~202502.
\newblock
  doi:{\changeurlcolor{black}\href{https://doi.org/10.1103/PhysRevLett.106.202502}{\detokenize{10.1103/PhysRevLett.106.202502}}}.

\bibitem[Pastore \em{et~al.}(2018)Pastore, Baroni, Carlson, Gandolfi, Pieper,
  Schiavilla, and Wiringa]{Pastore2018}
Pastore, S.; Baroni, A.; Carlson, J.; Gandolfi, S.; Pieper, S.C.; Schiavilla,
  R.; Wiringa, R.B.
\newblock {Quantum Monte Carlo calculations of weak transitions in A= 6–10
  nuclei}.
\newblock {\em Phys. Rev. C} {\bf 2018}, {\em 97},~022501,
  \href{http://xxx.lanl.gov/abs/1709.03592}{{\normalfont [1709.03592]}}.
\newblock
  doi:{\changeurlcolor{black}\href{https://doi.org/10.1103/PhysRevC.97.022501}{\detokenize{10.1103/PhysRevC.97.022501}}}.

\bibitem[Faestermann \em{et~al.}(2002)Faestermann, Schneider, Stolz,
  S{\"{u}}mmerer, Wafers, Friese, Geissel, Hellstr{\"{o}}m, Kienle,
  K{\"{o}}rner, Mineva, M{\"{u}}nch, M{\"{u}}nzenberg, Schlegel, Schmidt,
  Thirolf, Weick, and Zeitelhack]{Faestermann2002}
Faestermann, T.; Schneider, R.; Stolz, A.; S{\"{u}}mmerer, K.; Wafers, E.;
  Friese, J.; Geissel, H.; Hellstr{\"{o}}m, M.; Kienle, P.; K{\"{o}}rner, H.J.;
  Mineva, M.; M{\"{u}}nch, M.; M{\"{u}}nzenberg, G.; Schlegel, C.; Schmidt, K.;
  Thirolf, P.; Weick, H.; Zeitelhack, K.
\newblock {Decay studies of N $\approx$ Z nuclei from 75Sr to 102Sn}.
\newblock {\em Eur. Phys. J. A} {\bf 2002}, {\em 15},~185--188.
\newblock
  doi:{\changeurlcolor{black}\href{https://doi.org/10.1140/epja/i2001-10251-7}{\detokenize{10.1140/epja/i2001-10251-7}}}.

\bibitem[Batist \em{et~al.}(2010)Batist, G{\'{o}}rska, Grawe, Janas, Kavatsyuk,
  Karny, Kirchner, la~Commara, Mukha, Plochocki, and Roeckl]{Batist2010}
Batist, L.; G{\'{o}}rska, M.; Grawe, H.; Janas, Z.; Kavatsyuk, M.; Karny, M.;
  Kirchner, R.; la~Commara, M.; Mukha, I.; Plochocki, A.; Roeckl, E.
\newblock {Systematics of Gamow-Teller beta decay "Southeast" of 100Sn}.
\newblock {\em Eur. Phys. J. A} {\bf 2010}, {\em 46},~45--53.
\newblock
  doi:{\changeurlcolor{black}\href{https://doi.org/10.1140/epja/i2010-11025-x}{\detokenize{10.1140/epja/i2010-11025-x}}}.

\bibitem[Hinke \em{et~al.}(2012)Hinke, B{\"{o}}hmer, Boutachkov, Faestermann,
  Geissel, Gerl, Gernh{\"{a}}user, G{\'{o}}rska, Gottardo, Grawe,
  Gr{\c{e}}bosz, Kr{\"{u}}cken, Kurz, Liu, Maier, Nowacki, Pietri,
  Podoly{\'{a}}k, Sieja, Steiger, Straub, Weick, Wollersheim, Woods, Al-Dahan,
  Alkhomashi, Ata{\c{c}}, Blazhev, Braun, {\v{C}}elikovi{\'{c}}, Davinson,
  Dillmann, Domingo-Pardo, Doornenbal, de~France, Farrelly, Farinon, Goel,
  Habermann, Hoischen, Janik, Karny, Kaşkaş, Kojouharov, Kr{\"{o}}ll,
  Litvinov, Myalski, Nebel, Nishimura, Nociforo, Nyberg, Parikh,
  Proch{\'{a}}zka, Regan, Rigollet, Schaffner, Scheidenberger, Schwertel,
  S{\"{o}}derstr{\"{o}}m, Steer, Stolz, and Strmeň]{Hinke2012}
Hinke, C.B.; B{\"{o}}hmer, M.; Boutachkov, P.; Faestermann, T.; Geissel, H.;
  Gerl, J.; Gernh{\"{a}}user, R.; G{\'{o}}rska, M.; Gottardo, A.; Grawe, H.;
  Gr{\c{e}}bosz, J.L.; Kr{\"{u}}cken, R.; Kurz, N.; Liu, Z.; Maier, L.;
  Nowacki, F.; Pietri, S.; Podoly{\'{a}}k, Z.; Sieja, K.; Steiger, K.; Straub,
  K.; Weick, H.; Wollersheim, H.J.; Woods, P.J.; Al-Dahan, N.; Alkhomashi, N.;
  Ata{\c{c}}, A.; Blazhev, A.; Braun, N.F.; {\v{C}}elikovi{\'{c}}, I.T.;
  Davinson, T.; Dillmann, I.; Domingo-Pardo, C.; Doornenbal, P.C.; de~France,
  G.; Farrelly, G.F.; Farinon, F.; Goel, N.; Habermann, T.C.; Hoischen, R.;
  Janik, R.; Karny, M.; Kaşkaş, A.; Kojouharov, I.M.; Kr{\"{o}}ll, T.;
  Litvinov, Y.; Myalski, S.; Nebel, F.; Nishimura, S.; Nociforo, C.; Nyberg,
  J.; Parikh, A.R.; Proch{\'{a}}zka, A.; Regan, P.H.; Rigollet, C.; Schaffner,
  H.; Scheidenberger, C.; Schwertel, S.; S{\"{o}}derstr{\"{o}}m, P.A.; Steer,
  S.J.; Stolz, A.; Strmeň, P.
\newblock {Superallowed Gamow–Teller decay of the doubly magic nucleus
  100Sn}.
\newblock {\em Nature} {\bf 2012}, {\em 486},~341--345.
\newblock
  doi:{\changeurlcolor{black}\href{https://doi.org/10.1038/nature11116}{\detokenize{10.1038/nature11116}}}.

\bibitem[Lubos \em{et~al.}(2019)Lubos, Park, Faestermann, Gernh{\"{a}}user,
  Kr{\"{u}}cken, Lewitowicz, Nishimura, Sakurai, Ahn, Baba, Blank, Blazhev,
  Boutachkov, Browne, {\v{C}}elikovi{\'{c}}, {De France}, Doornenbal, Fang,
  Fukuda, Giovinazzo, Goel, G{\'{o}}rska, Ilieva, Inabe, Isobe, Jungclaus,
  Kameda, Kim, Kojouharov, Kubo, Kurz, Kwon, Lorusso, Moschner, Murai,
  Nishizuka, Patel, Rajabali, Rice, Schaffner, Shimizu, Sinclair,
  S{\"{o}}derstr{\"{o}}m, Steiger, Sumikama, Suzuki, Takeda, Wang, Warr,
  Watanabe, Wu, and Xu]{Lubos2019}
Lubos, D.; Park, J.; Faestermann, T.; Gernh{\"{a}}user, R.; Kr{\"{u}}cken, R.;
  Lewitowicz, M.; Nishimura, S.; Sakurai, H.; Ahn, D.S.; Baba, H.; Blank, B.;
  Blazhev, A.; Boutachkov, P.; Browne, F.; {\v{C}}elikovi{\'{c}}, I.; {De
  France}, G.; Doornenbal, P.; Fang, Y.; Fukuda, N.; Giovinazzo, J.; Goel, N.;
  G{\'{o}}rska, M.; Ilieva, S.; Inabe, N.; Isobe, T.; Jungclaus, A.; Kameda,
  D.; Kim, Y.K.; Kojouharov, I.; Kubo, T.; Kurz, N.; Kwon, Y.K.; Lorusso, G.;
  Moschner, K.; Murai, D.; Nishizuka, I.; Patel, Z.; Rajabali, M.M.; Rice, S.;
  Schaffner, H.; Shimizu, Y.; Sinclair, L.; S{\"{o}}derstr{\"{o}}m, P.A.;
  Steiger, K.; Sumikama, T.; Suzuki, H.; Takeda, H.; Wang, Z.; Warr, N.;
  Watanabe, H.; Wu, J.; Xu, Z.
\newblock {Improved Value for the Gamow-Teller Strength of the Sn 100 Beta
  Decay}.
\newblock {\em Phys. Rev. Lett.} {\bf 2019}, {\em 122},~222502.
\newblock
  doi:{\changeurlcolor{black}\href{https://doi.org/10.1103/PhysRevLett.122.222502}{\detokenize{10.1103/PhysRevLett.122.222502}}}.

\bibitem[Som{\`{a}} \em{et~al.}(2020)Som{\`{a}}, Navr{\'{a}}til, Raimondi,
  Barbieri, and Duguet]{Soma2020}
Som{\`{a}}, V.; Navr{\'{a}}til, P.; Raimondi, F.; Barbieri, C.; Duguet, T.
\newblock {Novel chiral Hamiltonian and observables in light and medium-mass
  nuclei}.
\newblock {\em Phys. Rev. C} {\bf 2020}, {\em 101},~014318,
  \href{http://xxx.lanl.gov/abs/1907.09790}{{\normalfont [1907.09790]}}.
\newblock
  doi:{\changeurlcolor{black}\href{https://doi.org/10.1103/PhysRevC.101.014318}{\detokenize{10.1103/PhysRevC.101.014318}}}.

\bibitem[Stroberg()]{imsrgcode}
Stroberg, S.R.
\newblock imsrg++ code.

\bibitem[Brown and Rae(2014)]{Brown2014}
Brown, B.A.; Rae, W.D.M.
\newblock {The Shell-Model Code NuShellX@MSU}.
\newblock {\em Nucl. Data Sheets} {\bf 2014}, {\em 120},~115--118.
\newblock
  doi:{\changeurlcolor{black}\href{https://doi.org/10.1016/j.nds.2014.07.022}{\detokenize{10.1016/j.nds.2014.07.022}}}.

\bibitem[Stroberg()]{nutbarcode}
Stroberg, S.R.
\newblock nutbar code.

\bibitem[Shimizu(2013)]{Shimizu2013}
Shimizu, N.
\newblock {Nuclear shell-model code for massive parallel computation, "KSHELL"}
  {\bf 2013}.
\newblock pp. 2--3,  \href{http://xxx.lanl.gov/abs/1310.5431}{{\normalfont
  [1310.5431]}}.

\bibitem[Shimizu \em{et~al.}(2019)Shimizu, Mizusaki, Utsuno, and
  Tsunoda]{Shimizu2019}
Shimizu, N.; Mizusaki, T.; Utsuno, Y.; Tsunoda, Y.
\newblock {Thick-restart block Lanczos method for large-scale shell-model
  calculations}.
\newblock {\em Comput. Phys. Commun.} {\bf 2019}, {\em 244},~372--384.
\newblock
  doi:{\changeurlcolor{black}\href{https://doi.org/10.1016/j.cpc.2019.06.011}{\detokenize{10.1016/j.cpc.2019.06.011}}}.

\bibitem[Bogner \em{et~al.}(2003)Bogner, Kuo, and Schwenk]{Bogner2003}
Bogner, S.K.; Kuo, T.T.S.; Schwenk, A.
\newblock {Model-independent low momentum nucleon interaction from phase shift
  equivalence}.
\newblock {\em Phys. Rep.} {\bf 2003}, {\em 386},~1--27.
\newblock
  doi:{\changeurlcolor{black}\href{https://doi.org/10.1016/j.physrep.2003.07.001}{\detokenize{10.1016/j.physrep.2003.07.001}}}.

\bibitem[Cohen and Kurath(1965)]{Cohen1965}
Cohen, S.; Kurath, D.
\newblock {Effective interactions for the 1p shell}.
\newblock {\em Nucl. Phys.} {\bf 1965}, {\em 73},~1--24.
\newblock
  doi:{\changeurlcolor{black}\href{https://doi.org/10.1016/0029-5582(65)90148-3}{\detokenize{10.1016/0029-5582(65)90148-3}}}.

\bibitem[Engel and Men{\'{e}}ndez(2017)]{Engel2017}
Engel, J.; Men{\'{e}}ndez, J.
\newblock {Status and Future of Nuclear Matrix Elements for Neutrinoless
  Double-Beta Decay: A Review}.
\newblock {\em Reports Prog. Phys.} {\bf 2017}, {\em 80},~046301,
  \href{http://xxx.lanl.gov/abs/1610.06548}{{\normalfont [1610.06548]}}.
\newblock
  doi:{\changeurlcolor{black}\href{https://doi.org/10.1088/1361-6633/aa5bc5}{\detokenize{10.1088/1361-6633/aa5bc5}}}.

\bibitem[Seng \em{et~al.}(2019)Seng, Gorchtein, and Ramsey-Musolf]{Seng2019}
Seng, C.Y.; Gorchtein, M.; Ramsey-Musolf, M.J.
\newblock { Dispersive evaluation of the inner radiative correction in neutron
  and nuclear $\beta$ decay }.
\newblock {\em Phys. Rev. D} {\bf 2019}, {\em 100},~13001,
  \href{http://xxx.lanl.gov/abs/1812.03352}{{\normalfont [1812.03352]}}.
\newblock
  doi:{\changeurlcolor{black}\href{https://doi.org/10.1103/physrevd.100.013001}{\detokenize{10.1103/physrevd.100.013001}}}.

\bibitem[Miller and Schwenk(2009)]{Miller2009}
Miller, G.A.; Schwenk, A.
\newblock {Isospin-symmetry-breaking corrections to superallowed Fermi $\beta$
  decay: Radial excitations}.
\newblock {\em Phys. Rev. C} {\bf 2009}, {\em 80},~064319.
\newblock
  doi:{\changeurlcolor{black}\href{https://doi.org/10.1103/PhysRevC.80.064319}{\detokenize{10.1103/PhysRevC.80.064319}}}.

\bibitem[Hebeler \em{et~al.}(2011)Hebeler, Bogner, Furnstahl, Nogga, and
  Schwenk]{Hebeler2011}
Hebeler, K.; Bogner, S.K.; Furnstahl, R.J.; Nogga, A.; Schwenk, A.
\newblock {Improved nuclear matter calculations from chiral low-momentum
  interactions}.
\newblock {\em Phys. Rev. C} {\bf 2011}, {\em 83},~031301.
\newblock
  doi:{\changeurlcolor{black}\href{https://doi.org/10.1103/PhysRevC.83.031301}{\detokenize{10.1103/PhysRevC.83.031301}}}.

\bibitem[Tichai \em{et~al.}(2019)Tichai, M{\"{u}}ller, Vobig, and
  Roth]{Tichai2019}
Tichai, A.; M{\"{u}}ller, J.; Vobig, K.; Roth, R.
\newblock {Natural orbitals for ab initio no-core shell model calculations}.
\newblock {\em Phys. Rev. C - Nucl. Phys.} {\bf 2019}, {\em 99},~034321,
  \href{http://xxx.lanl.gov/abs/1809.07571}{{\normalfont [1809.07571]}}.

\bibitem[Hoppe \em{et~al.}(2021)Hoppe, Tichai, Heinz, Hebeler, and
  Schwenk]{Hoppe2020}
Hoppe, J.; Tichai, A.; Heinz, M.; Hebeler, K.; Schwenk, A.
\newblock {Natural orbitals for many-body expansion methods}.
\newblock {\em Phys. Rev. C} {\bf 2021}, {\em 103},~014321,
  \href{http://xxx.lanl.gov/abs/2009.04701}{{\normalfont [2009.04701]}}.
\newblock
  doi:{\changeurlcolor{black}\href{https://doi.org/10.1103/PhysRevC.103.014321}{\detokenize{10.1103/PhysRevC.103.014321}}}.

\bibitem[Furnstahl \em{et~al.}(2012)Furnstahl, Hagen, and
  Papenbrock]{Furnstahl2012}
Furnstahl, R.J.; Hagen, G.; Papenbrock, T.
\newblock {Corrections to nuclear energies and radii in finite oscillator
  spaces}.
\newblock {\em Phys. Rev. C} {\bf 2012}, {\em 86},~031301.
\newblock
  doi:{\changeurlcolor{black}\href{https://doi.org/10.1103/PhysRevC.86.031301}{\detokenize{10.1103/PhysRevC.86.031301}}}.

\bibitem[Furnstahl \em{et~al.}(2014)Furnstahl, Phillips, and
  Wesolowski]{Furnstahl2015}
Furnstahl, R.J.; Phillips, D.R.; Wesolowski, S.
\newblock {A recipe for EFT uncertainty quantification in nuclear physics}.
\newblock {\em J. Phys. G Nucl. Part. Phys.} {\bf 2014}, {\em 42},~034028,
  \href{http://xxx.lanl.gov/abs/1407.0657}{{\normalfont [1407.0657]}}.
\newblock
  doi:{\changeurlcolor{black}\href{https://doi.org/10.1088/0954-3899/42/3/034028}{\detokenize{10.1088/0954-3899/42/3/034028}}}.

\bibitem[Hagen()]{HagenPC}
Hagen, G.
\newblock private communication.

\bibitem[Barrett \em{et~al.}(2013)Barrett, Navr{\'{a}}til, and
  Vary]{Barrett2013}
Barrett, B.R.; Navr{\'{a}}til, P.; Vary, J.P.
\newblock {Ab initio no core shell model}.
\newblock {\em Prog. Part. Nucl. Phys.} {\bf 2013}, {\em 69},~131--181.
\newblock
  doi:{\changeurlcolor{black}\href{https://doi.org/10.1016/j.ppnp.2012.10.003}{\detokenize{10.1016/j.ppnp.2012.10.003}}}.

\bibitem[Yao \em{et~al.}(2020)Yao, Bally, Engel, Wirth, Rodr{\'{i}}guez, and
  Hergert]{Yao2020}
Yao, J.M.; Bally, B.; Engel, J.; Wirth, R.; Rodr{\'{i}}guez, T.R.; Hergert, H.
\newblock {Ab Initio Treatment of Collective Correlations and the Neutrinoless
  Double Beta Decay of $^{48}$Ca}.
\newblock {\em Phys. Rev. Lett.} {\bf 2020}, {\em 124},~232501,
  \href{http://xxx.lanl.gov/abs/1908.05424}{{\normalfont [1908.05424]}}.
\newblock
  doi:{\changeurlcolor{black}\href{https://doi.org/10.1103/PhysRevLett.124.232501}{\detokenize{10.1103/PhysRevLett.124.232501}}}.

\bibitem[Belley \em{et~al.}(2021)Belley, Payne, Stroberg, Miyagi, and
  Holt]{Belley2020}
Belley, A.; Payne, C.G.; Stroberg, S.R.; Miyagi, T.; Holt, J.D.
\newblock {Ab Initio Neutrinoless Double-Beta Decay Matrix Elements for Ca 48 ,
  Ge 76 , and Se 82}.
\newblock {\em Phys. Rev. Lett.} {\bf 2021}, {\em 126},~042502,
  \href{http://xxx.lanl.gov/abs/2008.06588}{{\normalfont [2008.06588]}}.
\newblock
  doi:{\changeurlcolor{black}\href{https://doi.org/10.1103/PhysRevLett.126.042502}{\detokenize{10.1103/PhysRevLett.126.042502}}}.

\bibitem[Yao \em{et~al.}(2021)Yao, Belley, Wirth, Miyagi, Payne, Stroberg,
  Hergert, and Holt]{Yao2021}
Yao, J.M.; Belley, A.; Wirth, R.; Miyagi, T.; Payne, C.G.; Stroberg, S.R.;
  Hergert, H.; Holt, J.D.
\newblock {Ab initio benchmarks of neutrinoless double-$\beta$ decay in light
  nuclei with a chiral Hamiltonian}.
\newblock {\em Phys. Rev. C} {\bf 2021}, {\em 103},~014315,
  \href{http://xxx.lanl.gov/abs/2010.08609}{{\normalfont [2010.08609]}}.
\newblock
  doi:{\changeurlcolor{black}\href{https://doi.org/10.1103/PhysRevC.103.014315}{\detokenize{10.1103/PhysRevC.103.014315}}}.

\bibitem[Hagen \em{et~al.}(2010)Hagen, Papenbrock, and
  Hjorth-Jensen]{Hagen2010b}
Hagen, G.; Papenbrock, T.; Hjorth-Jensen, M.
\newblock {Ab initio computation of the F17 proton halo state and resonances in
  A=17 nuclei}.
\newblock {\em Phys. Rev. Lett.} {\bf 2010}, {\em 104},~5--8,
  \href{http://xxx.lanl.gov/abs/1003.1995}{{\normalfont [1003.1995]}}.
\newblock
  doi:{\changeurlcolor{black}\href{https://doi.org/10.1103/PhysRevLett.104.182501}{\detokenize{10.1103/PhysRevLett.104.182501}}}.

\bibitem[Heinz \em{et~al.}(2021)Heinz, Tichai, Hoppe, Hebeler, and
  Schwenk]{Heinz2021}
Heinz, M.; Tichai, A.; Hoppe, J.; Hebeler, K.; Schwenk, A.
\newblock {In-medium similarity renormalization group with three-body
  operators}.
\newblock {\em Phys. Rev. C} {\bf 2021}, {\em 103},~044318,
  \href{http://xxx.lanl.gov/abs/2102.11172}{{\normalfont [2102.11172]}}.
\newblock
  doi:{\changeurlcolor{black}\href{https://doi.org/10.1103/PhysRevC.103.044318}{\detokenize{10.1103/PhysRevC.103.044318}}}.

\bibitem[Simonis \em{et~al.}(2017)Simonis, Stroberg, Hebeler, Holt, and
  Schwenk]{Simonis2017}
Simonis, J.; Stroberg, S.R.; Hebeler, K.; Holt, J.D.; Schwenk, A.
\newblock {Saturation with chiral interactions and consequences for finite
  nuclei}.
\newblock {\em Phys. Rev. C} {\bf 2017}, {\em 96},~014303,
  \href{http://xxx.lanl.gov/abs/1704.02915}{{\normalfont [1704.02915]}}.
\newblock
  doi:{\changeurlcolor{black}\href{https://doi.org/10.1103/PhysRevC.96.014303}{\detokenize{10.1103/PhysRevC.96.014303}}}.

\bibitem[Taniuchi \em{et~al.}(2019)Taniuchi, Santamaria, Doornenbal, Obertelli,
  Yoneda, Authelet, Baba, Calvet, Ch{\^{a}}teau, Corsi, Delbart, Gheller,
  Gillibert, Holt, Isobe, Lapoux, Matsushita, Men{\'{e}}ndez, Momiyama,
  Motobayashi, Niikura, Nowacki, Ogata, Otsu, Otsuka, P{\'{e}}ron, P{\'{e}}ru,
  Peyaud, Pollacco, Poves, Rouss{\'{e}}, Sakurai, Schwenk, Shiga, Simonis,
  Stroberg, Takeuchi, Tsunoda, Uesaka, Wang, Browne, Chung, Dombradi, Franchoo,
  Giacoppo, Gottardo, Hady{\'{n}}ska-Kl{\c{e}}k, Korkulu, Koyama, Kubota, Lee,
  Lettmann, Louchart, Lozeva, Matsui, Miyazaki, Nishimura, Olivier, Ota, Patel,
  Şahin, Shand, S{\"{o}}derstr{\"{o}}m, Stefan, Steppenbeck, Sumikama, Suzuki,
  Vajta, Werner, Wu, and Xu]{Taniuchi2019}
Taniuchi, R.; Santamaria, C.; Doornenbal, P.; Obertelli, A.; Yoneda, K.;
  Authelet, G.; Baba, H.; Calvet, D.; Ch{\^{a}}teau, F.; Corsi, A.; Delbart,
  A.; Gheller, J.M.; Gillibert, A.; Holt, J.D.; Isobe, T.; Lapoux, V.;
  Matsushita, M.; Men{\'{e}}ndez, J.; Momiyama, S.; Motobayashi, T.; Niikura,
  M.; Nowacki, F.; Ogata, K.; Otsu, H.; Otsuka, T.; P{\'{e}}ron, C.;
  P{\'{e}}ru, S.; Peyaud, A.; Pollacco, E.C.; Poves, A.; Rouss{\'{e}}, J.Y.;
  Sakurai, H.; Schwenk, A.; Shiga, Y.; Simonis, J.; Stroberg, S.R.; Takeuchi,
  S.; Tsunoda, Y.; Uesaka, T.; Wang, H.; Browne, F.; Chung, L.X.; Dombradi, Z.;
  Franchoo, S.; Giacoppo, F.; Gottardo, A.; Hady{\'{n}}ska-Kl{\c{e}}k, K.;
  Korkulu, Z.; Koyama, S.; Kubota, Y.; Lee, J.; Lettmann, M.; Louchart, C.;
  Lozeva, R.; Matsui, K.; Miyazaki, T.; Nishimura, S.; Olivier, L.; Ota, S.;
  Patel, Z.; Şahin, E.; Shand, C.; S{\"{o}}derstr{\"{o}}m, P.A.; Stefan, I.;
  Steppenbeck, D.; Sumikama, T.; Suzuki, D.; Vajta, Z.; Werner, V.; Wu, J.; Xu,
  Z.Y.
\newblock {78Ni revealed as a doubly magic stronghold against nuclear
  deformation}.
\newblock {\em Nature} {\bf 2019}, {\em 569},~53--58.
\newblock
  doi:{\changeurlcolor{black}\href{https://doi.org/10.1038/s41586-019-1155-x}{\detokenize{10.1038/s41586-019-1155-x}}}.

\bibitem[Stroberg \em{et~al.}(2021)Stroberg, Holt, Schwenk, and
  Simonis]{Stroberg2021}
Stroberg, S.R.; Holt, J.D.; Schwenk, A.; Simonis, J.
\newblock {Ab initio Limits of Atomic Nuclei}.
\newblock {\em Phys. Rev. Lett.} {\bf 2021}, {\em 126},~022501,
  \href{http://xxx.lanl.gov/abs/1905.10475}{{\normalfont [1905.10475]}}.
\newblock
  doi:{\changeurlcolor{black}\href{https://doi.org/10.1103/PhysRevLett.126.022501}{\detokenize{10.1103/PhysRevLett.126.022501}}}.

\bibitem[Hagen \em{et~al.}(2016)Hagen, Jansen, and Papenbrock]{Hagen2016}
Hagen, G.; Jansen, G.R.; Papenbrock, T.
\newblock {Structure of Ni 78 from First-Principles Computations}.
\newblock {\em Phys. Rev. Lett.} {\bf 2016}, {\em 117},~172501.
\newblock
  doi:{\changeurlcolor{black}\href{https://doi.org/10.1103/PhysRevLett.117.172501}{\detokenize{10.1103/PhysRevLett.117.172501}}}.

\bibitem[Gaarde(1983)]{Gaarde1983}
Gaarde, C.
\newblock {Gamow-Teller and M1 resonances}.
\newblock {\em Nucl. Physics, Sect. A} {\bf 1983}, {\em 396},~127--144.
\newblock
  doi:{\changeurlcolor{black}\href{https://doi.org/10.1016/0375-9474(83)90014-3}{\detokenize{10.1016/0375-9474(83)90014-3}}}.

\end{thebibliography}

%

\end{document}